\documentclass[sn-mathphys-num]{sn-jnl}

\usepackage{graphicx}%
\usepackage{multirow}%
\usepackage{amsmath,amssymb,amsfonts}%
\usepackage{amsthm}%
\usepackage{mathrsfs}%
\usepackage[title]{appendix}%
\usepackage{xcolor}%
\usepackage{textcomp}%
\usepackage{manyfoot}%
\usepackage{booktabs}%
\usepackage{algorithm}%
\usepackage{algorithmicx}%
\usepackage{algpseudocode}%
\usepackage{listings}%

\usepackage{hyperref}

\let\orcidlogo\relax
\makeatother
\usepackage{orcidlink}

\usepackage{subcaption}

\usepackage{rotating}

\theoremstyle{thmstyleone}%
\theoremstyle{thmstyletwo}%

\theoremstyle{thmstylethree}%

\begin{document}

\title{Potential Landscapes Reveal Spatiotemporal Structure in Urban Mobility: Hodge Decomposition and Principal Component Analysis of Tokyo Before and During COVID-19}


\author[1]{\fnm{Yunhan} \sur{Du}\orcidlink{0000-0002-1785-6919}}\email{du.yunhan.work@gmail.com}

\author[2]{\fnm{Takaaki} \sur{Aoki}\orcidlink{0000-0002-3312-3061}}\email{takaaki.aoki.work@gmail.com}

\author*[1,3,4,5]{\fnm{Naoya} \sur{Fujiwara}\orcidlink{0000-0002-7753-4698}}\email{naoya.fujiwara@tohoku.ac.jp}

\affil[1]{\orgdiv{Graduate School of Information Sciences}, \orgname{Tohoku University}, \orgaddress{\street{6-3-09 Aoba, Aramaki, Aoba}, \city{Sendai}, \postcode{980-8579}, \state{Miyagi}, \country{Japan}}}

\affil[2]{\orgdiv{Graduate School of Data Science}, \orgname{Shiga University}, \orgaddress{\street{1-1-1, Banba}, \city{Hikone}, \postcode{522-8522}, \state{Shiga}, \country{Japan}}}

\affil[3]{\orgdiv{PRESTO}, \orgname{Japan Science and Technology Agency}, \orgaddress{\street{4-1-8 Honcho}, \city{Kawaguchi}, \postcode{332‑0012}, \state{Saitama}, \country{Japan}}}

\affil[4]{\orgdiv{Center for Spatial Information Science}, \orgname{The University of Tokyo}, \orgaddress{\street{5-1-5 Kashiwanoha}, \city{Kashiwa}, \postcode{277‑8568}, \state{Chiba}, \country{Japan}}}

\affil[5]{\orgdiv{Institute of Industrial Science}, \orgname{The University of Tokyo}, \orgaddress{\street{4-6-1 Komaba}, \city{Meguro-ku}, \postcode{153‑8505}, \state{Tokyo}, \country{Japan}}}

\footnotetext{Google scholar profiles: Yunhan Du \url{https://scholar.google.com/citations?user=zgv6kzkAAAAJ&hl=en}; Takaaki Aoki \url{https://scholar.google.com/citations?user=e8x3baAAAAAJ&hl=en}; Naoya Fujiwara \url{https://scholar.google.co.jp/citations?user=qmt1jPMAAAAJ&hl=en}}


\abstract{
Understanding human mobility is vital to solving societal challenges, such as epidemic control and urban transportation optimization.
Recent advancements in data collection now enable the exploration of dynamic mobility patterns in human flow.
However, the vast volume and complexity of mobility data make it difficult to interpret spatiotemporal patterns directly, necessitating effective information reduction.
The core challenge is to balance data simplification with information preservation: methods must retain location-specific information about human flows from origins to destinations while reducing the data to a comprehensible level.
This study proposes a two-step dimensionality reduction framework: First, combinatorial Hodge theory is applied to the given origin--destination (OD) matrices with timestamps to construct a set of potential landscapes of human flow, preserving imbalanced trip information between locations.
Second, principal component analysis (PCA) expresses the time series of potential landscapes as a linear combination of a few static spatial components, with their coefficients representing temporal variations.
The framework systematically decouples the spatial and temporal components of the given data.
By implementing this two-step reduction method, we reveal large weight variations during a pandemic, characterized by an overall decline in mobility and stark contrasts between weekdays and holidays.
These findings demonstrate the effectiveness of our framework in uncovering complex mobility patterns and its potential to inform urban planning and public health interventions.}




\maketitle

\section{Introduction}\label{sec_introduction}
Humans are inherently mobile beings who frequently travel from one location to another for various daily activities, such as commuting, shopping, entertainment, and education. These individual movements collectively shape the intricate dynamics of urban mobility.
Determining the spatiotemporal patterns of human flow are crucial for a wide range of applications, including controlling epidemics spread \cite{JLY2020, VBS2006, HSA2021}, understanding social and economic issues such as social segregation \cite{AKK2014, MCD2021, YPM2023} and urban gender disparities \cite{GTP2020, CDG2023}, and integrating public transportation systems with urban design  to enhance mobility efficiency \cite{VSR2018, XCG2021, BZK2021, GPB2016}.

Human flow is not static but fluctuates dynamically over time. Therefore, understanding the evolution of these patterns is crucial. For example, on weekdays, cities exhibit distinct human flow patterns during morning and evening rush hours, whereas on weekends, movement trends differ significantly. Furthermore, external factors such as natural disasters and pandemics can drastically alter these patterns. Natural disasters, such as earthquakes or floods, can abruptly disrupt natural human movement, triggering mass evacuations or substantial changes in commuting behaviors. Similarly, the outbreak of a pandemic, such as Coronavirus Disease 2019 (COVID-19) \cite{ZZW2020}, can cause unprecedented shifts in human mobility, with lockdowns and travel restrictions drastically reducing movement and reshaping human flow patterns on a global scale. These temporal variations in human flow result from a complex interplay of factors, including economic activity, social events, environmental conditions, and policy interventions. Examining these patterns over time provides valuable insights into the underlying mechanisms driving human mobility.

Recent advancements in data collection and processing technologies have enabled the analysis of large-scale mobility data at high spatiotemporal resolutions, facilitating the exploration of dynamic mobility patterns and the identification of underlying structures in human flow. We live in an era where vast amounts of human mobility data are generated from various sources \cite{BDK2015, Z2015, ZTL2016, ZYW2018, YJR2022}. Traditionally, mobility data have been gathered through censuses and surveys that collect information on where people live and work. With the increasing prevalence of mobile phones, call detail records (CDRs) have become an important data source. These records, provided by mobile network operators, track the location of a signal tower when a phone call or text message is sent. Additionally, global positioning system (GPS) data collected from GPS-enabled devices and smartphones offers increasingly detailed location information. Social media platforms also contribute valuable mobility data through user interactions and geotagged posts.

Although vast amounts of mobility data are now available, their sheer volume and complexity make it difficult to understand the spatiotemporal patterns of human flow on the Earth's surface, necessitating effective information reduction methods to avoid information overload.
Table \ref{tbl:comparison_mobility} summarizes the information reduction methods in previous studies.
In earlier research, mobility data have often been reduced to temporal trends of macroscopic statistics, such as displacement and waiting time distributions \cite{BHG2006}, radius of gyration \cite{GHB2008}, number of visited places and visitation frequency \cite{SKW2010}, predictability \cite{SQB2010}, motifs \cite{SBC2013}, and mobility synchronization \cite{SBB2023}. Some studies have examined the dynamics of one or more of these metrics \cite{KSZ2020, HSU2013, LBH2012, GY2017, WT2016, HRH2020, KSH2022, SBB2023}, revealing common patterns in human mobility. Nevertheless, these data processing approaches often mitigate information overload by reducing detail, resulting in the loss of location-specific information, which prevents discussion of mobility patterns at specific locations on a map.
In other instances, data have been transformed into spatial distributions of temporal populations \cite{BLT2011, LXM2016, LST2017}, preserving location-specific details but losing crucial mobility information, such as trip distances and their origins. Additionally, data have been used to fit general models of human mobility, including the gravity model \cite{Z1946}, intervening opportunities model \cite{S1940, FM1990, AP2001, NSL2012, SYB2015}, radiation model \cite{SGM2012, LZD2013, YZF2014, REW2014, KLG2015, VTN2018, LY2019, LY2020, DF2022}, and exploration and preferential return (EPR) model \cite{PSR2015, SDO2021, BLE2015, ASS2018, LZZ2021}. 
Although these models offer valuable insights, the fitted parameters are usually macroscopic,
and location-specific information is often lost.
In summary, the core challenge is to reconcile data simplification with information preservation: To study mobility patterns on a map, retaining location-specific information about human flows from origins to destinations is crucial. At the same time, the data must be simplified sufficiently to be comprehensible. 

\begin{sidewaystable}
\centering
\caption{Comparison of information reduction methods of human mobility data}
\label{tbl:comparison_mobility}
\begin{tabular}{p{4cm}p{4cm}p{4cm}p{4cm}}
\toprule
Method & Captured properties & Advantages & Limitations \\
\midrule
Fit information to general models of human mobility  (e.g., gravity, radiation, intervening opportunities, EPR models) \cite{Z1946,S1940, FM1990, AP2001, NSL2012, SYB2015,SGM2012, LZD2013, YZF2014, REW2014, KLG2015, VTN2018, LY2019, LY2020, DF2022, PSR2015, SDO2021, BLE2015, ASS2018, LZZ2021}   & 
Calibrated model parameters to given datasets &
Abstract real-world patterns to fit theoretical constructs. &
Often loss of location-specific information if the model has no such information\\
\midrule
Reduce information to temporal trends of macroscopic statistics \cite{BHG2006, GHB2008, SKW2010, SQB2010, SBC2013, SBB2023, KSZ2020, HSU2013, LBH2012, GY2017, WT2016, HRH2020, KSH2022} &
Temporal trends of macroscopic statistics  (e.g., displacement, radius of gyration, predictability, motifs, synchronization) &
Effectively reduces the complexity and scale of mobility data, and is helpful in revealing common patterns in humans mobility and examining the dynamics of metrics over time.&
Loss of location-specific information \\
\midrule
Transform information into spatial distributions of temporal populations \cite{BLT2011, LXM2016, LST2017} & 
Time series of population distributions &
Preserves location-specific details of population distribution&
Loss of crucial mobility information, such as trip distances and their origins \\
\midrule
Two-step dimensionality reduction (Our method) & 
Potential landscape as static spatial components and a trajectory of time-varying coefficients of these components &
Preserves essential location-specific spatial patterns of human mobility and their temporal changes &
Loss of circular components of human flow, which is orthogonal to the gradient component derived from potential\\
\bottomrule
\end{tabular}
\end{sidewaystable}

\begin{figure}[h] 
  \centering 
  \includegraphics[width=\textwidth]{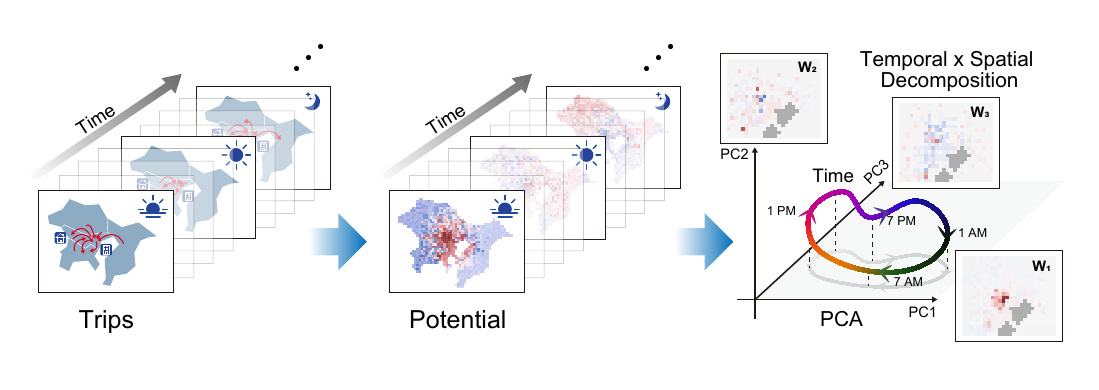}
  \caption{Methodology for examining spatiotemporal mobility patterns. (Left) Raw, hourly OD data derived from mobility datasets serve as the foundation of the analysis, capturing detailed flow patterns between spatial grids. (Middle) Hourly OD data transformed into a series of hourly potential landscapes. (Right) Principal component axes representing the contribution of each spatial grid to the corresponding eigenvector $\mathbf{w}_{k}$. Hourly dynamics are defined by the trajectory of score changes over time.}\label{fig_schematic}
\end{figure}

Therefore, our goal is to balance simplicity with information fidelity to study human flow on a map using information reduction methods.
This study proposes a two-step reduction method for human mobility data, enabling the effective representation of spatiotemporal patterns in human flow.
Fig.~\ref{fig_schematic} illustrates the concept of our method.
Let us consider a human flow data as a time series of origin--destination (OD) matrices (left panel in Fig.~\ref{fig_schematic}).
The first reduction step involves utilizing the concept of the scalar potential \cite{AFF2022, AFF2023, MMB2019, SOT2022, YLG2024}.
In our case, the potential is generated through applying the combinatorial Hodge theory \cite{K1949, H1989, JLY2011, BBS2025} to the OD matrix at each timestamp. 
This method has key advantages.
(1) It retains location-specific information about human flows from origins to destinations.
Higher-potential locations collect the people's movements from lower-potential locations.
It extracts the imbalanced component of a given flow, which can be expressed by the differences of potentials between locations (see Fig.~\ref{fig:hodge} and Methods for details).
(2) It offers a significant reduction in data dimensionality.
An OD matrix represents the trips from origins to destinations.
It signifies relational data between locations (i.e., locations $\times$ locations dimensionality), and is not easily shown on a map.
By reducing the OD matrix to the potential at each location, the obtained potential landscape is a location-level statistic, which can be naturally visualized on a map.  
After the first step, we get a time series of potential landscapes (middle panel in Fig.~\ref{fig_schematic}), like time slices of a movie of a map.
The second step decouples the temporal and spatial components.
Using principal component analysis (PCA), the time series of potential landscapes can be approximated by linear combinations of a few potential landscapes.
The potential landscapes (eigenvector $\mathbf{w}_{k}$) are static and only retain spatial information of human mobility.
The mixing of these potential landscapes changes with time and the coefficients (principal component score) retain the temporal information of human mobility.
As a result of the two-step reduction, spatiotemporal behavior in human flow can be represented as a trajectory in a low-dimensional space, as shown in the right panel of Fig.~\ref{fig_schematic}.
Compared with the raw OD matrices, this two-step reduction generates a small number of potential landscapes (corresponding to the PCA axes) that provide reduced, location-specific flow information. In contrast, the resulting trajectory offers a clear representation of temporal changes. 

As a case study, we apply the reduction method to study a dataset in Tokyo before and during COVID-19.
Do the first dominant principal components capture weekday/holiday and pandemic temporal signatures?
The findings reveal significant shifts in mobility behavior during the pandemic, including decreased overall movement and high spatially constrained mobility patterns, particularly during peak commute hours.
Additionally, a clear difference is observed between weekday and holiday mobility, with holidays showing delayed peaks and less spatial dispersion, both before and during the pandemic.

The remainder of this paper is organized as follows: Section~\ref{sec_data} describes the mobility dataset, which contains hourly OD information for both pre-COVID-19 and COVID-19 periods. Section~\ref{sec_methods} provides an overview of combinatorial Hodge theory (Section~\ref{sec_methods_hodge}) and explains its application to human flow data (Section~\ref{method_application_hodge}).
In Section~\ref{sec_results}, we begin with the extraction of the scalar potential field of human flow through the unique decomposition of a given OD matrix using combinatorial Hodge theory. Then, PCA is applied to the potential landscape to investigate temporal variations in human flows within 24 hours across different scenarios, including weekday versus weekend trips and pre-COVID-19 versus during COVID-19 periods. 
Finally, Section~\ref{sec_discussion} summarizes our key findings and concludes the paper. These results highlight the profound impact of COVID-19 on urban mobility and provide insights into human flow adaptations to external disruptions, thereby offering valuable implications for urban planning and public health policies.

\section{Data}\label{sec_data}
For this study, we use Japanese human mobility data, ``LocationMind xPop,'' provided by LocationMind Inc. \cite{LocationMind}, that encompasses GPS records collected within the Tokyo metropolitan area. The dataset spans two time periods: May and June 2019 (pre-COVID-19) and May and June 2021 (during COVID-19). These data include estimated home locations, origins, and destinations, which are mapped using a 2km by 2km spatial grid system based on the Japanese mesh index \cite{MESH}. The dataset distinguishes between weekday and holiday movements and records the specific hours during which these movements occur. Notably, the data are aggregated, reflecting average flows over multiple days rather than individual trips. Additionally, the number of trips is based on the unique users recorded, with adjustments against population census totals, providing an estimate of real population movements.

The ``LocationMind xPop'' data refer to human flow data derived from individual location records transmitted by smartphones with users' consent, provided by NTT DOCOMO, INC.
To ensure privacy, the data provider processed the data in aggregate using statistical methods, protecting any personally identifiable information.
The original location data comprised GPS coordinates (latitude and longitude) recorded at intervals as short as 5 minutes and did not include any information that could directly identify individuals.

We outline several limitations of the dataset. First, we relied on aggregated two-month mobility averages for 2019 and 2021, which limited detection of transient fluctuations, such as those driven by weather, short-term events, or sudden policy shifts. Second, we adjusted the GPS data using population census data, assuming proportional relationships between the sampled device users and the broader population. This assumption may overlook variations in local population density and disparities in device usage, thereby introducing bias.

The background map layouts in Figs.~\ref{fig_potential_dynamics} and \ref{fig_timeseries} are based on data from  \textcopyright{} OpenStreetMap contributors, available under the Open Database License (ODbL). Map tiles are provided by \textcopyright{} CARTO. The results in Figs.~\ref{fig_potential_dynamics}, \ref{fig_timeseries}, \ref{fig_pca_comparison_holiday}, and \ref{fig_pca_comparison_COVID} were obtained by using data from ``LocationMind xPop \textcopyright{} LocationMind Inc'' \cite{LocationMind}.

\section{Methods}\label{sec_methods}
\subsection{Combinatorial Hodge theory}
\label{sec_methods_hodge}

\begin{figure}[h] 
  \centering 
  \includegraphics[width=\textwidth]{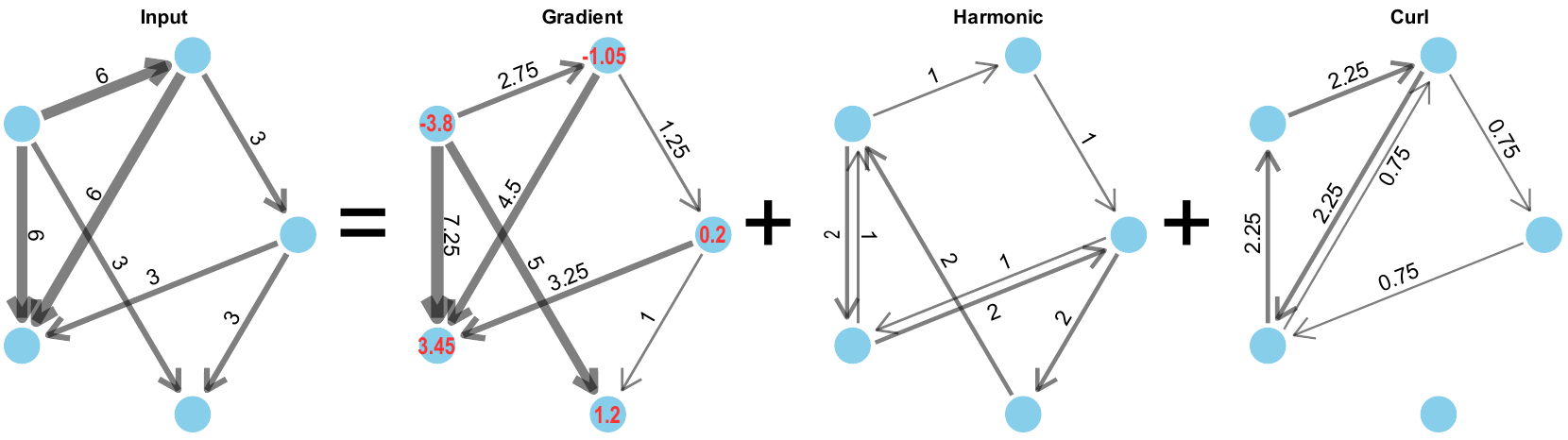}
  \caption{Illustration of potential extraction by combinatorial Hodge decomposition. Input edge flow $Y$ in the left panel can be decomposed into three orthogonal components: gradient, harmonic, and curl flows. 
The gradient flow is the focal component of this reduction method, which can be described by differences in node potentials, denoted by red numbers.
The other elements are cyclic, and incoming and outgoing fluxes are balanced, while the gradient component represents the imbalanced flow that induces the temporal changes of population mass.
}
\label{fig:hodge}
\end{figure}

As the first step of reduction, we apply combinatorial Hodge theory \cite{JLY2011}, a mathematical framework that extracts the gradient component of a given flow from the other cyclic components. The gradient component represents the imbalanced flow of the given flow that induces the temporal changes of population mass: it can be derived from the differences of ``potentials'' at each location, and we refer to the geospatial distribution of the potentials as potential landscape \cite{AFF2022, AFF2023}.
We favor Hodge gradients because they summarize net source–sink structure interpretable on maps.
This potential landscape provides a dimensionality reduction of human flow.
The human flow---the trips from origins to destinations---represents relational data between locations (i.e., locations $\times$ locations dimensionality), and the reduced potential landscape is a location-level statistic.


To describe the potential landscape mathematically,
we consider an undirected graph $G(V,E)$, where $V$ denotes a set of vertices (corresponding to locations) and $E$ represents a set of edges. Edge flow $Y$ is assigned to each edge, with each element $Y_{ij}$ representing the flow from vertex $i$ to vertex $j$. The edge flow is skew-symmetric, i.e., $Y_{ij} = -Y_{ji}$. The combinatorial gradient, curl, and divergence are defined as follows \cite{JLY2011}:
\begin{align}
  (\text{grad}\, s)(i, j) &=  s_j-s_i \quad \text{for $\{i,j\} \in E$} , \\
(\text{curl}\, Y)(i, j, k) &= Y_{ij} + Y_{jk} + Y_{ki}\quad \text{for  $\{i,j,k\}$}: \{i, j\}, \{j, k\}, \{k, i\} \in E ,\\
  (\text{div}Y)(i) &= \sum_{j \text{ s.t. } \{i,j\} \in E} Y_{ij},
\end{align}
where $s$ denotes the potential to be introduced.

The space of edge flow $\mathcal{Y}$ is orthogonally decomposed into the images and kernels of these operators:
\begin{equation} 
  \mathcal{Y}  = \text{im}(\text{grad})  \oplus \text{ker}(\Delta_1) \oplus  \text{im}(\text{curl}^*), \label{eq:decomposition}
\end{equation}
where ker($\Delta_1$) = ker(curl) $\cap$ ker(div) and $\text{curl}^*$ is the adjoint operator of  curl.

Potential $s$ is obtained by solving  the weighted least-squares optimization problem given by
\begin{align} \label{eq:optimization}
  \min_s \sum_{\{i,j\} \in E} W_{ij} \left[ \text{(grad $s$)}(i,j) - Y_{ij} \right]^2
  =
  \min_s \sum_{\{i,j\} \in E} W_{ij} \left[ (s_j - s_i) - Y_{ij} \right]^2,
\end{align}
where $W_{ij} (\in [0,1])$ is the weight for each pairwise comparison, primarily determined by the distance $d_{ij}$ as discussed in Section~\ref{method_application_hodge}. The optimization problem determines the closest point to the given data $Y$ in the subspace of edge flows, which comprises the gradient flows of the potential. The problem can be solved by an $l_{2}$-projection of $Y$ onto $\text{im}(\text{grad})$. With a Euclidean inner product in space $\mathcal{Y}$, $\langle X,Y\rangle = \sum_{ \{i,j\} \in E} W_{ij} X_{ij}Y_{ij}$, the normal equation can be expressed as:
\begin{align}
  \Delta_0 s  = - \text{div} Y,  \label{eq:normal_equation}
\end{align}
where $\Delta_0$ is the graph Laplacian given by
\begin{align}
  \left[ \Delta_{0} \right]_{ij} =  \begin{cases}
    \sum_j W_{ij} \quad &\text{if $i = j$}\\
    -W_{ij}    \quad &\text{if $\{i,j\} \in E$}\\
    0    \quad &\text{otherwise}.
  \end{cases}
\end{align}

Finally, the potential $s$ is given by the minimal-norm solution of Eq.~\ref{eq:normal_equation} as:
\begin{align}
  s  = - \Delta_0^{\dagger} \text{div} Y,  \label{eq:potential}
\end{align}
where $\dagger$ denotes the Moore--Penrose inverse.

\subsection{Application of combinatorial Hodge theory to mobility flow} \label{method_application_hodge}
For a given OD matrix $M$ representing the mobility flow, we assign edge flow $Y$ by calculating the net flow as:
\begin{align}
  Y = M - M^{\top},
t\end{align}
where the element $Y_{ij}$ indicates the net flow from $i$ to $j$, satisfying the necessary skew-symmetric condition in the combinatorial Hodge theory \cite{K1949, H1989, JLY2011}. 
To define the potential using the optimization problem in Eq.~\ref{eq:optimization}, the regional difference in potential, $s_{j}-s_{i}$, is compared with  the net flow $Y_{ij}$, but only for location pairs included in the edge set $E$. This evaluation is further weighted using $W_{ij}$, which encodes the importance  of each pair in the analysis.

The determination of edges $E$ and weights $W_{ij}$ depends primary on the distances between locations. To explore the dependence of mobility flow on distance, we analyze the share of nonzero mobility pairs ($M_{ij} > 0$ or $M_{ji} > 0$) as a function of road distance, using data from the Japanese mobility dataset (Section~\ref{sec_data}). The results appear in the Supplementary Information. Our analysis reveals a sharp decline in the percentage of non-zero mobility pairs as the distance increases. Notably, even at a relatively short distance of approximately $d_{ij} \sim 40$ km, nearly half of the mobility pairs exhibit zero flow volume. 

The phenomenon of zero-mobility flow admits several interpretations.
A possible explanation is that mobility incurs substantial costs, and zero flow suggests that the costs associated with travel outweigh the drivers of regional differences.
If the high costs prevent many people from traveling,
the absence of mobility flow conveys little useful information about regional potential differences, and it will be better to ignore it in the optimization by \eqref{eq:optimization}.
At shorter distances, where travel costs will be generally low, zero mobility flow between two locations implies that the regional difference $(s_{j}-s_{i})$ is too minimal to induce travel. In this situation, the absence of mobility flow provides valuable information and should be considered in the optimization in \eqref{eq:optimization}.
Importantly, while we cannot identify the specific reasons preventing travel, if trips are consistently unobserved, this zero-flow pair conveys no information about regional potential differences and is therefore excluded from the analysis.

\begin{figure}[h!] 
  \centering
  \includegraphics[width=\textwidth]{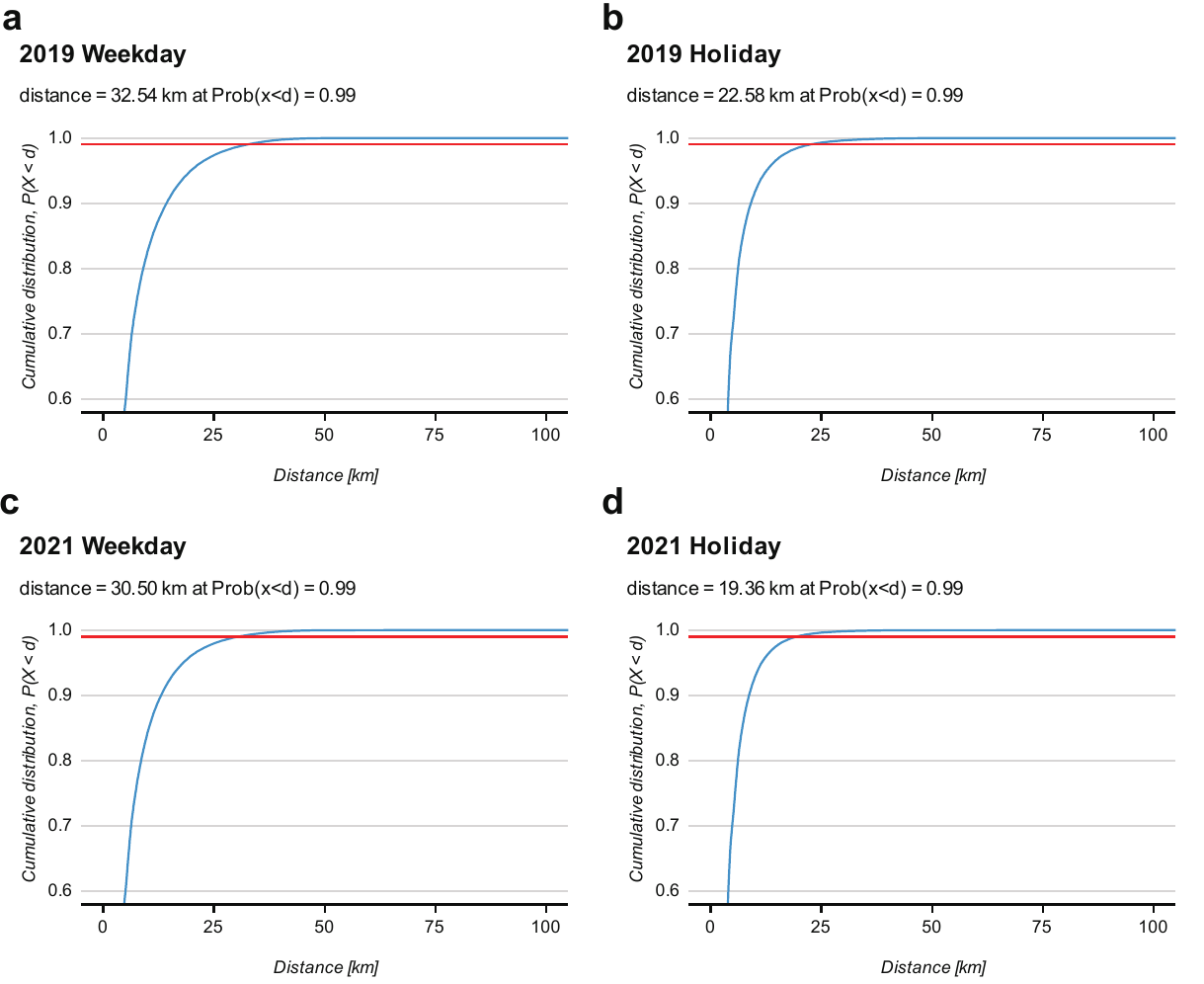}
  \caption{Threshold distances $\theta$ across four scenarios: (a) 2019 Weekday; (b) 2019 Holiday; (c) 2021 Weekday; and (d) 2021 Holiday.}
  \label{fig_threshold}
\end{figure}

To exclude the zero mobility flow with little information,
we apply a weighting method with a threshold distance $\theta$, which has been shown to perform well in urban settings \cite{AFF2023}:
\begin{align}
  W_{ij} = \begin{cases}
    1 \quad  &\text{mobility pair $\{ (i,j) \mid d_{ij} \le \theta\}$} \\
    0 \quad & \text{mobility pair $\{ (i,j) \mid d_{ij} > \theta\}$}
\end{cases}.
\label{eq:weight}
\end{align}
The threshold $\theta$ is the distance below which the majority of trips occur, using the probabilistic criterion $p(X < \theta) = 0.99$ (Fig.~\ref{fig_threshold}).
The value ($p$ = 0.99) is chosen to include the majority of trips while excluding possible outliers with less than one percent.
We confirmed that the obtained results by this specific criterion with $p = 0.99$ are robust against small perturbations on $p$ (see Supplementary Information for details).
For the set of edges $E$, we retain pairs with positive weights, $\{ (i,j) \mid W_{ij} =1 \}$.

\subsection{Principal component analysis (PCA)}
\label{method_pca}

PCA is a statistical technique commonly used for dimensionality reduction in data analysis. Previous research has used PCA to analyze and interpret human mobility patterns \cite{SYW2011, CBT2013, SCL2021, LGL2021}.
It transforms a high-dimensional dataset into a lower-dimensional set of orthogonal components that capture the maximum variance in the original data. In this study, we seek to characterize patterns in human mobility data by applying PCA to the potential landscape of human flow. Our objective is to identify the placement of spatial grids in the potential landscape and examine how the principal components contribute to the temporal evolution of human flow patterns.

Consider a dataset represented by a matrix $\mathbf{X}$ with $n$ rows and $p$ columns.
Each row $r$, denoted as $\mathbf{x}^{(r)}$, is a vector representing a single observation with $p$ features.
In this case, $\mathbf{X}$ represents a series of potentials across spatial grids, where $p$ is the number of spatial grids, and $n$ is the number of observations under various conditions.
Specifically, $n=96$ covers 24 hours across different scenarios: weekdays in 2019, holidays in 2019, weekdays in 2021, and holidays in 2021. For example, a particular column might represent the potential at 08:00 on weekdays in 2019.
$\mathbf{X}$ has zero column-wise means because the potentials given by Eq.~\ref{eq:potential} sum to zero, and the data were not standardized.

The primary goal of PCA is to identify features that explain the majority of the variance in the data.
First, we compute the covariance matrix $\mathbf{Q}$ of the dataset $\mathbf{X}$ as:
\begin{equation}
\mathbf{Q} = \frac{1}{n-1} \sum_{r=1}^n (\mathbf{x}^{(r)})^{\top} \mathbf{x}^{(r)} = \frac{1}{n-1} \mathbf{X}^{\top}\mathbf{X}.
\end{equation}
Performing eigenvalue decomposition on $\mathbf{Q}$ yields a set of eigenvalues $\lambda_{1}, \lambda_{2}, \cdots, \lambda_{p}$ and their corresponding eigenvectors $\mathbf{w}_{1}, \mathbf{w}_{2}, \cdots, \mathbf{w}_{p}$. The eigenvectors $\mathbf{w}_{k}$ represent the directions in the feature space along which the data vary the most. Meanwhile, the eigenvalues $\lambda_{k}$ indicate the amount of variance captured by each corresponding eigenvector $\mathbf{w}_{k}$. The transformation is then performed as follows:
\begin{equation}
\mathrm{PC}_{k}^{(r)} = \mathbf{x}^{(r)} \cdot \mathbf{w}_{k} \quad \text{for} \quad r = 1, \dots, n \quad \text{and} \quad k = 1, \dots, l,
\end{equation}
where $\mathrm{PC}_{k}^{(r)}$ is defined as the $k-$th principal component \textit{score} of the data point in row $r$. For this study, we specifically denote these scores as PC1, PC2, PC3, and so on.
Notably, the original data points $\mathbf{x}^{(r)}$ can be reconstructed from these principal components as follows:
\begin{equation}
    \mathbf{x}^{(r)} = \sum_{k=1}^{l}\text{PC}_k^{(r)}\mathbf{w}_k,
\end{equation}
which clearly shows PCA as a linear transformation that decomposes the mobility data into temporal components $\text{PC}_k^{(r)}$ and spatial patterns $\mathbf{w}_k$.

The $k$-th eigenvalue $\lambda_k$ equals the sum of the squares of the scores over $n$ observations. It can be expressed as:
\begin{equation}
\lambda_k = \sum_{i=1}^n (\mathrm{PC}_{k}^{(r)})^2 = \sum_{i=1}^n (\mathbf{x}^{(r)} \cdot \mathbf{w}_k)^2.
\end{equation}

In this study, we examine the eigenvectors and principal component scores to interpret the outcomes of PCA and their implications for human mobility patterns. As illustrated in Fig.~\ref{fig_schematic} (right panel), PCA decomposes the potential landscape into separable spatial and temporal dimensions. The axes PC1, PC2, and PC3 represent the spatial contributions of each principal direction, denoted by the eigenvectors $\mathbf{w}_k$. The temporal evolution of mobility is captured through color-coded trajectories that depict the hourly variation in the principal component scores.

\section{Results}\label{sec_results}

\subsection{Temporal evolution of potential landscape}\label{results_1} 
\begin{figure}[h] 
  \centering 
  \includegraphics[width= \textwidth]{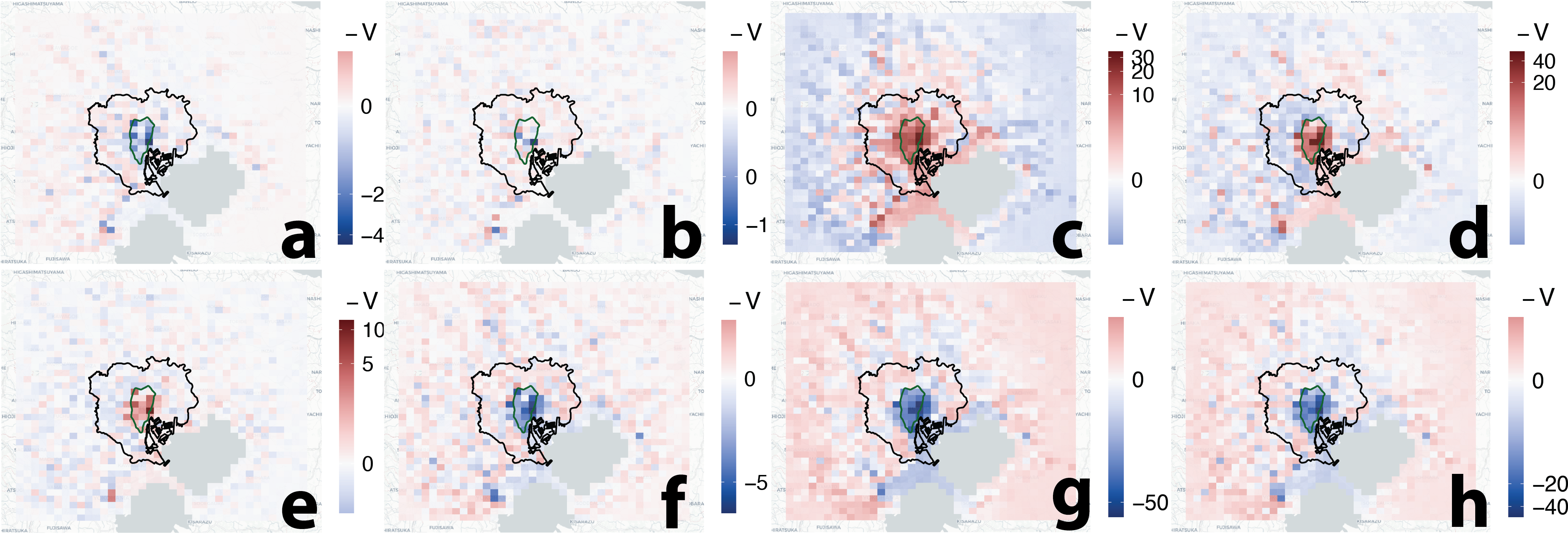}
  \caption{Temporal evolution of the potential landscape $-V(=s)$ for weekdays in 2019 in the Tokyo metropolitan area. Grid colors represent the potential value in each spatial grid. Black contour lines delineate the borders of Tokyo's special wards, marking the primary urban center. Green contour lines denote the Yamanote railway loop, which encloses the urban core of Tokyo. Subfigures how different hours: (a) 00:00; (b) 03:00; (c) 06:00; (d) 09:00; (e) 12:00; (f) 15:00; (g) 18:00; (h) 21:00.}\label{fig_potential_dynamics} 
\end{figure}

Using Japanese human mobility data (Section~\ref{sec_data}), we initially compute the potential landscape of human flow at different hours of the day. The results for weekdays in 2019 are presented in Fig.~\ref{fig_potential_dynamics}, displaying the spatial landscape at 3-h intervals (hourly details are provided in the Supplementary Information). In these maps, the Yamanote railway loop is represented by a green line enclosing the Tokyo urban core, linking major centers such as Shinjuku, Shibuya, Ikebukuro, Ueno, and Tokyo Station. Areas beyond the loop mainly comprise suburban regions and nearby cities.

During the morning rush hours, from 06:00 (Fig.~\ref{fig_potential_dynamics}c) to 09:00 (Fig.~\ref{fig_potential_dynamics}d), the potential values in the central area were strongly positive, whereas several grids in the peripheral areas exhibited negative potential values. This pattern indicated the significant movement of estimated commuters from residential zones to the central business district, where a high concentration of workplaces is located. The strong central pull observed during these hours aligned with typical commuting patterns.

In the evening hours at approximately 18:00 (Fig.~\ref{fig_potential_dynamics}g), the peripheral areas showed elevated levels, while the urban core displayed negative values. 
This pattern indicated the end of the workday and the outward movement of individuals from central workplaces as they returned to their residential areas. The trend persisted into the late evening (Fig.~\ref{fig_potential_dynamics}h), with negative potential values remaining in the central areas and positive values dominating in the periphery. These trends also captured post-work activities, such as social gatherings, before individuals returned home.
These findings align with well-established commuting dynamics in urban environments, highlighting the utility of potential landscapes in capturing fundamental patterns of human mobility.

\subsection{Spatiotemporal dynamics revealed through PCA on potential landscape}\label{results_2} 
\begin{figure}[h] 
  \centering 
  \includegraphics[width=\textwidth]{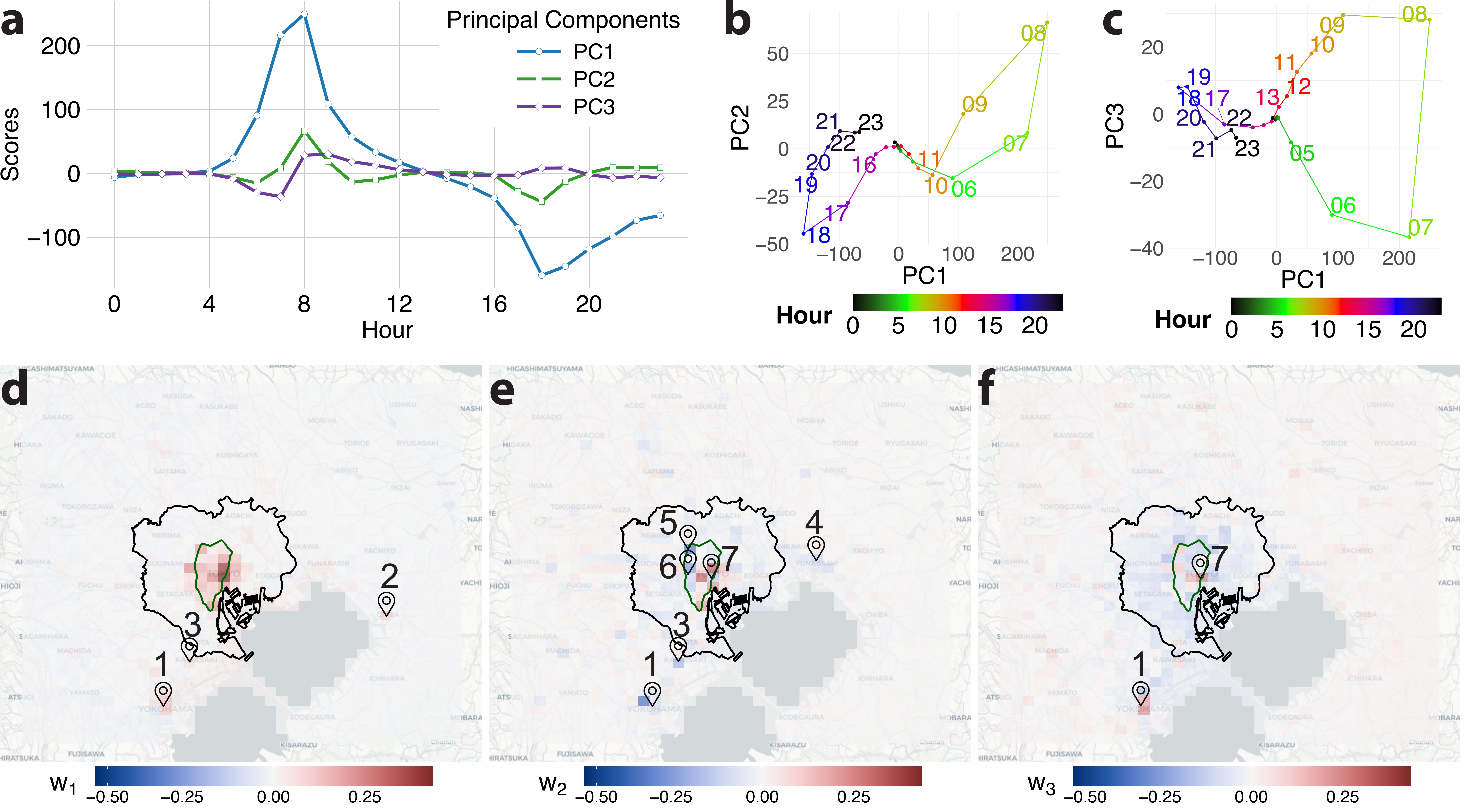}
  \caption{Temporal evolution of the PC1, PC2, and PC3 scores, and their corresponding eigenvectors on weekdays in 2019. (a) Time series of the principal component scores over a day. (b) and (c) Relations between principal components in PC2 vs. PC1 and PC3 vs. PC1 spaces, respectively. The corresponding eigenvectors are shown in (d) $\mathbf{w}_{1}$, (e) $\mathbf{w}_{2}$, and (f) $\mathbf{w}_{3}$. Numbers on the maps (d)--(f) indicate key locations: 1-Yokohama, 2-Chiba, 3-Kawasaki, 4-Funabashi, 5-Ikebukuro, 6-Shinjuku, and 7-Tokyo Station. Contour lines mirror those in Fig.~\ref{fig_potential_dynamics}.}\label{fig_timeseries} 
\end{figure}

Additionally, we performed PCA on the potential landscapes for different scenarios, including weekdays and holidays in 2019 and 2021 (Section~\ref{method_pca}). Notably, the first three principal components accounted for 96.3\% of the variance (Supplementary Information), highlighting their crucial role in explaining the underlying mobility patterns. By focusing on these dominant components, we mapped the temporal sequence of potential landscapes into a trajectory within the three-dimensional space of the principal components. As illustrated in the right panel of Fig.~\ref{fig_schematic}, this trajectory captured temporal information, while the corresponding eigenvectors $\mathbf{w}_{k}$ conveyed the spatial characteristics of the flow.
These eigenvectors are the static potential landscapes, and their linear combinations with temporal coefficients approximate the temporal dynamics.
The results for weekdays in 2019 appear in Fig.~\ref{fig_timeseries}. 

\begin{figure}[h] 
  \centering 
  \includegraphics[width=\textwidth]{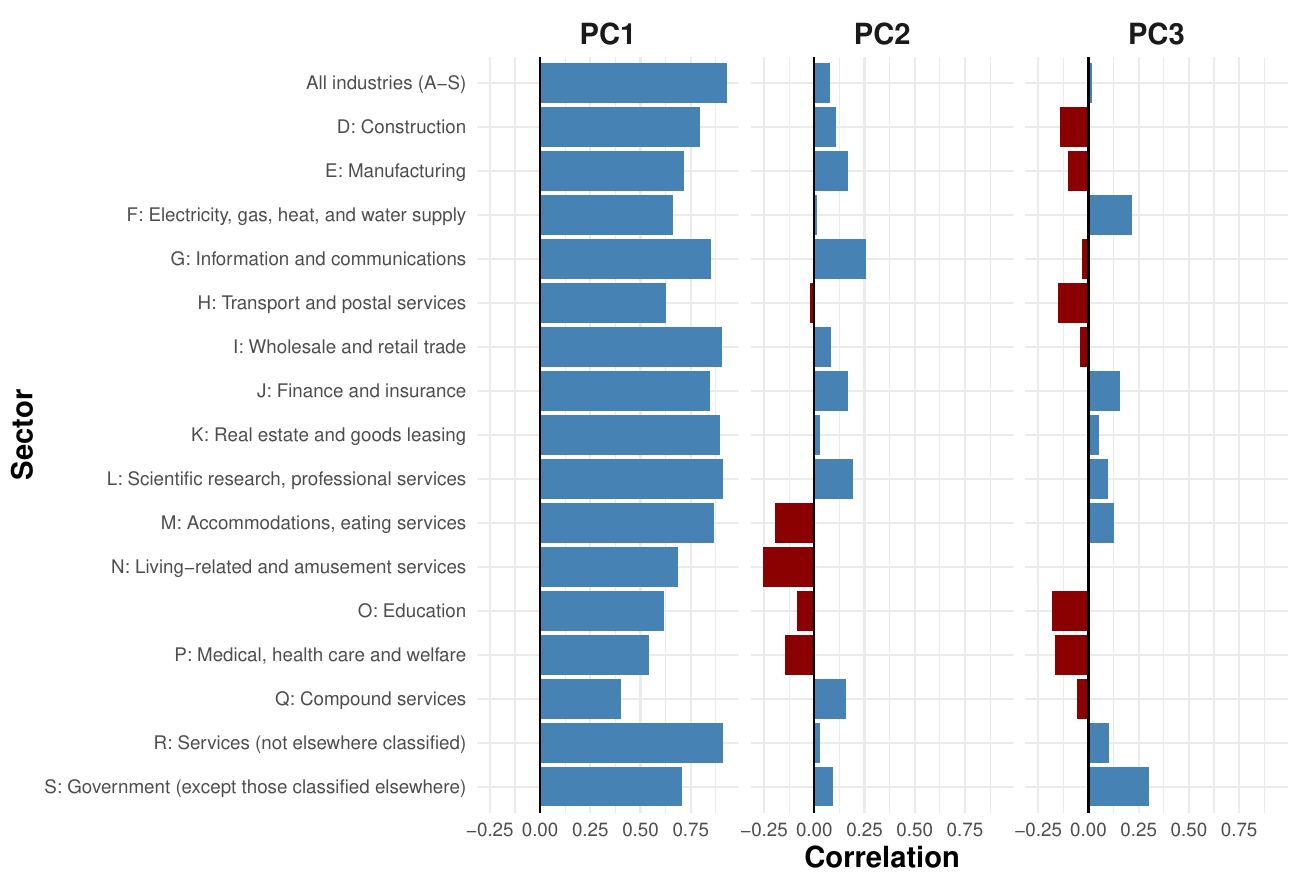}
  \caption{Correlations between the spatial characteristics of the principal components (eigenvectors $\mathbf{w}_1$, $\mathbf{w}_2$, $\mathbf{w}_3$) and spatial distribution of employees by industrial sectors.}
  \label{fig_eigen} 
\end{figure}

PC1 primarily captured the dominant commuting patterns.
Temporally, its score peaks during the morning rush hours (06:00--09:00) and dips in the evening (17:00--22:00), as shown in Fig.~\ref{fig_timeseries}a.
Spatially, the potential landscapes from its corresponding eigenvector $\mathbf{w}_1$ (Fig.~\ref{fig_timeseries}d) were concentrated within the special wards of Tokyo, particularly inside the Yamanote loop line, which constituted the central business district and extended to several key subcenters of the metropolis, including Chiba, Kawasaki, and Yokohama.
This spatial distribution of potentials, combined with the temporal trend, indicated the typical flow of estimated commuters from suburban regions into urban centers in the morning, followed by return trips in the evening. Notably, PC1 consistently exhibited higher scores than PC2 and PC3, highlighting its primary role in capturing dominant mobility dynamics.

This interpretation finds support in the pearson-correlation analysis between the eigenvector $\mathbf{w}_1$ and the number of employees by industry (Fig.~\ref{fig_eigen}). 
The data on employees are drawn from the 2021 Economic Census for Business Activity conducted by the Statistics Bureau of Japan, and we used the major category defined by the Japan Standard Industrial Classification.
Notably, the employees in the Agriculture and Forestry and Fisheries sectors are not individually shown in the data, and the Mining sector is omitted in this analysis because employment is minimal compared with other sectors.
The eigenvector $\mathbf{w}_1$ of PC1 was positively correlated with the number of employees in every industry and showed a strong correlation (0.933 $\pm$ 0.007) with the total number of employees (all industries).
This observation indicates that PC1 primarily captures the major commuting trips as the basic pattern of the daily mobility.

PC2 captured commuting behavior but exhibited more localized spatiotemporal characteristics.
Its score peaked at approximately 08:00 and dropped between 17:00 and 18:00 (Fig.~\ref{fig_timeseries}a).
The eigenvector $\mathbf{w}_{2}$ (Fig.~\ref{fig_timeseries}e) was positive near Tokyo Station and Shinagawa Station—both major high-speed rail hubs—and negatively concentrated around terminal stations, such as Ikebukuro, Shinjuku, Kawasaki, Funabashi, and Yokohama. These terminals are located along multiple radial lines extending from Tokyo Station.
As shown in Fig.~\ref{fig_eigen}, $\mathbf{w}_2$ of PC2 had a weak positive correlation ($\sim$0.25) with ``G: Information and Communications,'' whereas it shows a weak negative correlation ($\sim$ -0.25) with ``N: Living-related and Amusement Services'' and ``M: Accommodations, Eating and Drinking Services.''
This implies that PC2-driven flow is industry-related.
Consequently, PC2 primarily captured industry-related flows from these subcenter regions toward central high-speed rail hubs in the morning, followed by return trips in the early evening.

By contrast, PC3 exhibited a distinct temporal pattern, declining from 06:00 to 07:00 before sharply increasing between 08:00 and 10:00 (Fig.~\ref{fig_timeseries}a).
The eigenvector $\mathbf{w}_{3}$ (Fig.~\ref{fig_timeseries}f) was predominantly positive in suburban areas and negative in central districts, with isolated positive hotspots near Tokyo and Yokohama Stations.
$\mathbf{w}_3$ of PC3 had a weak positive correlation with ``S: Government (except those classified elsewhere)'' and ``F: Electricity, Gas, Heat, and Water Supply,'' whereas it shows a weak negative correlation with ``O: Education and P: Medical, Health Care and Welfare.''
These findings suggest that PC3 served as a complementary modulation to the commuting flows captured by PC1 and PC2.
Early morning with a negative coefficient to $\mathbf{w}_3$, the employees in the education and medical sectors commute from suburban areas to central districts,
and later. the employees in the government and public utility sectors move toward railway nodes.
Furthermore, PC3 shows smoother yet more frequent variations throughout the day, capturing midday activity and smaller-scale movements, such as work-related travel, thereby contributing to the overall daily mobility patterns.

\subsection{Comparison of human mobility patterns across scenarios}\label{results_3} 
Compared with the time series of potential landscapes, the PCA-based representation offered a macroscopic view of spatiotemporal patterns through a compact set of principal components. As shown in Fig.~\ref{fig_timeseries}, the trajectory in principal component space encoded the temporal signatures of these patterns. We compared the trajectories across different conditions—weekdays vs. weekends and 2019 vs. 2021—within the same three-dimensional space, enabling a systematic comparison of mobility trends.
\begin{figure}[h] 
  \centering 
  \includegraphics[width=\textwidth]{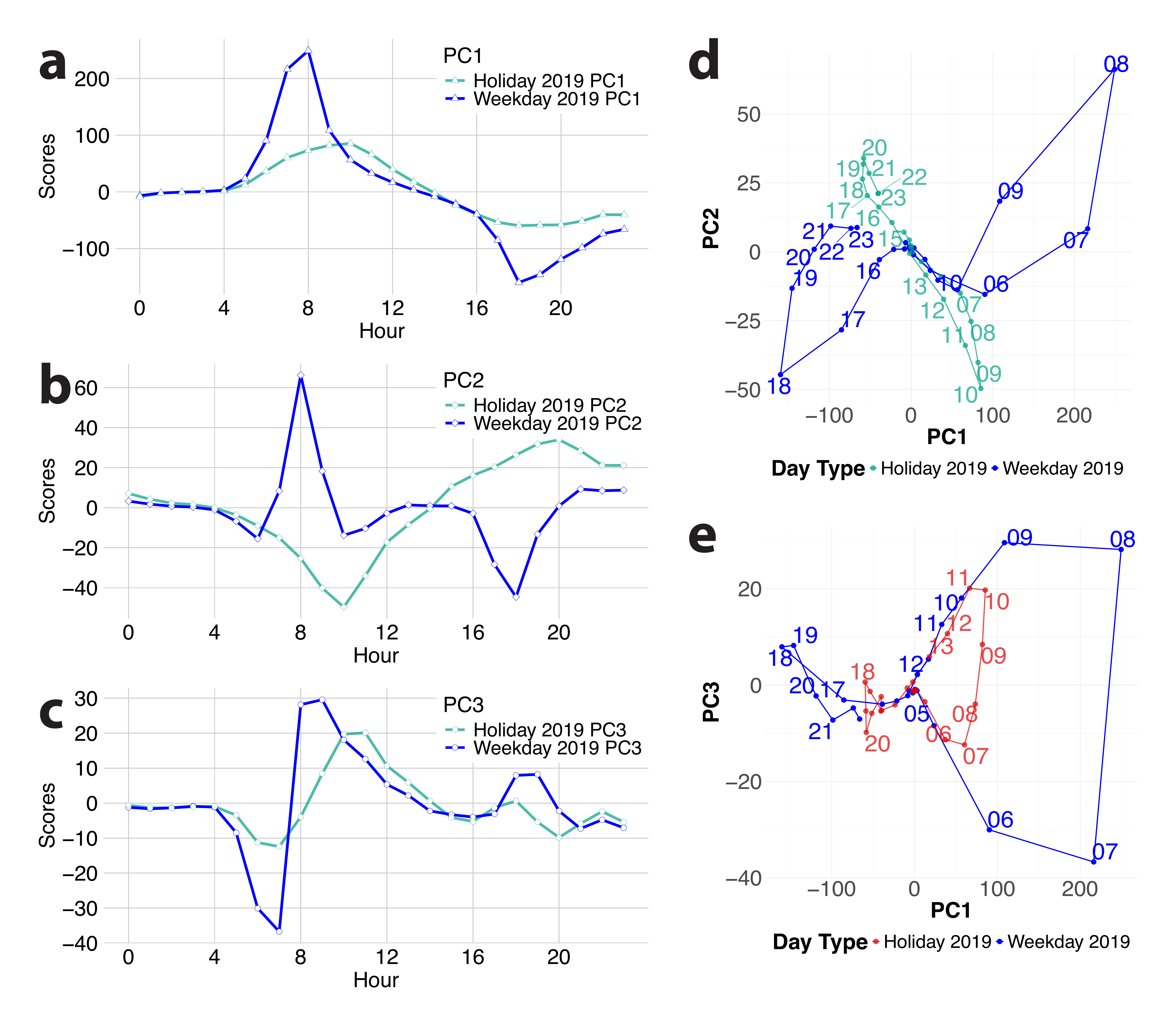}
  \caption{Comparison of weekday and holiday trajectories for 2019 in the principal component space. (a)--(c) Temporal evolution of PC1, PC2, and PC3 scores, respectively. (d) Trajectories projected onto the PC1-PC2 plane. (e) Trajectories projected onto the PC1-PC3 plane. }\label{fig_pca_comparison_holiday} 
\end{figure}

The weekday and holiday score trajectories for 2019 appear in Fig.~\ref{fig_pca_comparison_holiday}, highlighting distinct temporal signatures. As discussed in Section~\ref{results_2}, PC1 and PC3 encode weekday commuting flows and complementary movements. On holidays, PC1 followed a similar trajectory to weekdays, as shown in Fig.~\ref{fig_pca_comparison_holiday}a, but with a lower overall amplitude and a peak occurring 1 hour later. The scores of PC3 displayed a comparable pattern, as shown in Fig.~\ref{fig_pca_comparison_holiday}c. These findings suggest that commuting-like travel persisted on holidays but with reduced intensity and a later start compared to weekdays.

PC2 exhibited substantially different behaviors on holidays than on weekdays. As shown in Fig.~\ref{fig_pca_comparison_holiday}b, the PC2 score was negative on holidays in the morning (07:00--12:00) and positive in the evening (16:00--23:00), whereas on weekdays, the PC2 score was positive in the morning and negative in the evening. This opposite trend in PC2 scores on holidays captured flows originating from central high-speed rail stations toward suburban terminals along radial lines, reversing the weekday flow pattern.
As shown in Fig.~\ref{fig_eigen}, $\mathbf{w}_2$ of PC2 had a weak negative correlation ($\sim$ -0.25) with ``N: Living-related and Amusement Services'' and ``M: Accommodations, Eating and Drinking Services.''
Combined with its negative scores on holidays, this flow may represent leisure and entertainment trips.
Moreover, Fig.~\ref{fig_pca_comparison_holiday}b shows pronounced temporal differences in PC2 between weekdays and holidays. On weekdays, PC2 peaked sharply at approximately 08:00, while on holidays, it exhibited a broad dip from 07:00 to 12:00, reaching a minimum near 10:00. A similar shift appeared in the evening, with extended activity periods. These findings suggest that on holidays, PC2 primarily reflected leisure and entertainment trips rather than weekday commuting flows.
\begin{figure}[h] 
  \centering 
  \includegraphics[width=\textwidth]{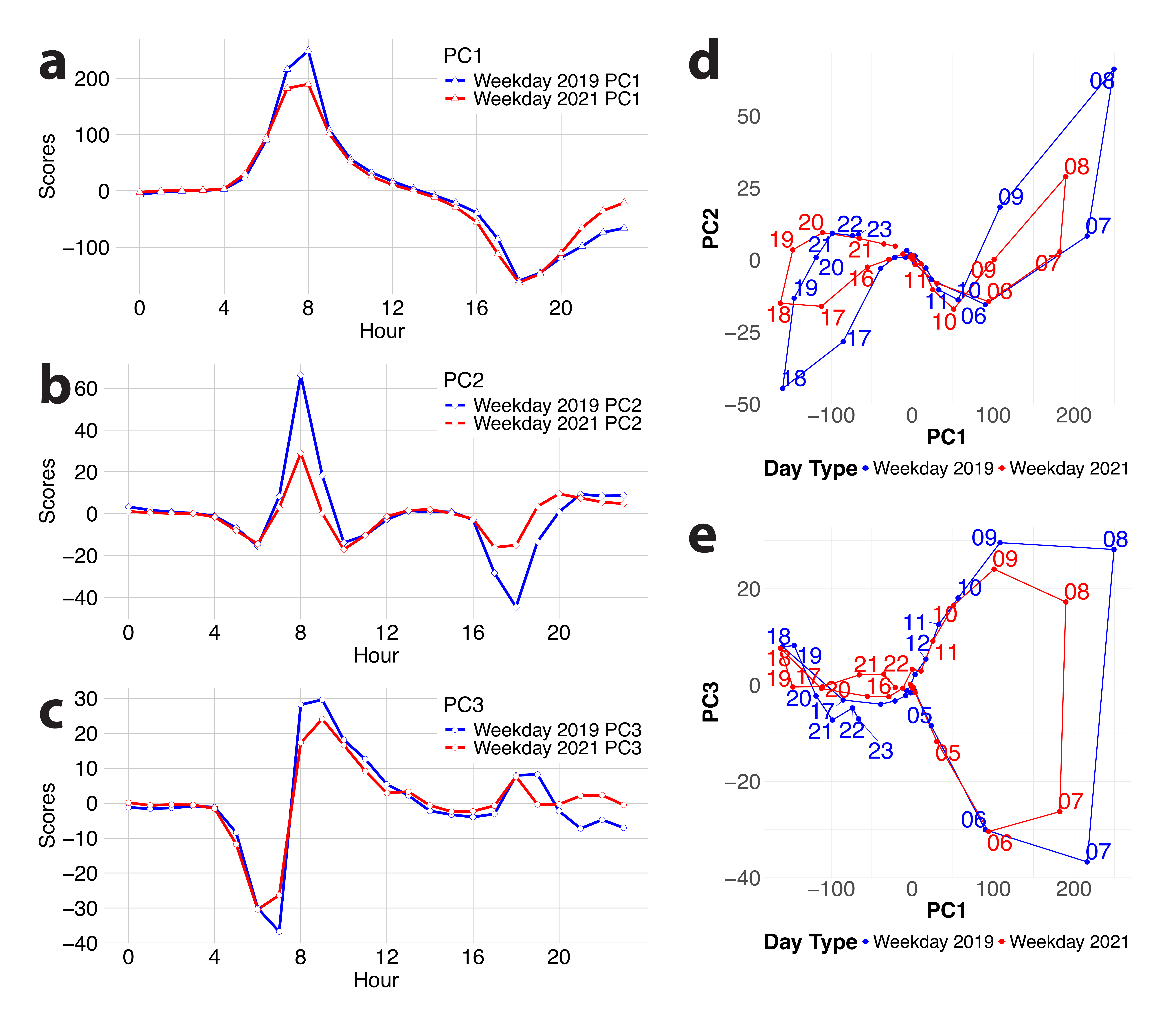}
  \caption{Comparison of 2019 and 2021 trajectories on weekdays in the principal component space. (a)--(c) Temporal evolution of PC1, PC2, and PC3 scores, respectively. (d) Trajectories projected onto the PC1-PC2 plane. (e) Trajectories projected onto the PC1-PC3 plane. }\label{fig_pca_comparison_COVID} 
\end{figure}

Additionally, we compared mobility trajectories before and during the COVID-19 pandemic. The weekday trajectories for 2019 and 2021 appear in Fig.~\ref{fig_pca_comparison_COVID}.
Although both trajectories had a comparable form---indicating that the fundamental structure of mobility patterns remained largely unchanged---the 2021 trajectory exhibited noticeably lower amplitudes during peak hours. For example, morning peaks in PC1 and PC2 were smaller (Fig.~\ref{fig_pca_comparison_COVID}a--c), and the evening return-trip dip in PC2 was also reduced and ended sooner. By contrast, the amplitude differences at other times of the day were less pronounced.

During the data collection period (May and June 2021), the Japanese government declared an emergency in Tokyo to curb the COVID-19 resurgence ahead of the 2021 Tokyo Summer Olympics. Although mobility restrictions in Japan were relatively lenient compared to international measures, prior research reported an overall decline in metropolitan mobility \cite{KNR2022, HA2022}. Our findings Additionally show that these restrictions primarily reduced mobility during peak commuting hours, whereas daily movement patterns outside these periods remained relatively unchanged.

\section{Discussion}\label{sec_discussion}
This study systematically analyzes human mobility patterns in the Tokyo metropolitan area using a proposed analytical framework based on potential landscape and PCA.
By comparing behaviors across distinct contexts, we identify salient differences in movement dynamics.
Our method effectively captures differences in daily mobility patterns between weekdays and holidays and detects substantial shifts in mobility behavior during the COVID-19 pandemic.
These findings elucidate how human mobility adapts to external factors, such as public health crises, across both temporal and spatial dimensions.
Notably, the proposed approach uses the potential landscape framework to represent and analyze the dynamics of human flow.
This method serves as an effective tool by capturing the temporal evolution of spatial potential distributions. Mapping these shifts onto a dynamic landscape reveals fine-grained variations in behavior, such as constrained mobility during peak hours in 2021 and notable differences in holiday and weekday movements.
Unlike static mobility data, the proposed approach offers deeper insight into the response of human movements to routine activities and external disruptions.

Nevertheless, this study has several limitations. First, while the proposed reduction method is suitable for generic timestamped origin--destination datasets, we have only studied the Tokyo dataset before and during COVID-19. Applications to other cities or other contexts remain for future work. Second, notably, the potential landscape derived by the combinatorial Hodge theory does not capture the information of cycle flows between locations. As illustrated in Fig.~\ref{fig:hodge}, this reduction extracts the gradient component of a given human flow while omitting the other cyclic components.
Unlike the gradient component, such cyclic components do not induce temporal changes in the local population, because the incoming and outgoing fluxes are balanced at each place.
In other words, the proposed method only focuses on the imbalanced population flows toward higher-potential places from lower-potential places.
Third, the proposed method cannot directly identify trip purpose or activity types from the data. For example, while we infer that specific patterns represent commuting, this is an inference based on spatiotemporal evidence. This issue warrants future studies. One approach is to leverage person-trip survey data with travel purpose or activity types. Another solution is integrating GPS-tracked mobility data with external point-of-interest datasets to infer the activity more directly.

Nonetheless, the proposed method offers a macroscopic view of spatiotemporal mobility patterns, rendering it well-suited for applications that do not require a high level of detail.
Our findings have direct relevance for several applications.
A deeper understanding of mobility patterns can inform urban planning, support the design of efficient transportation networks, and guide the allocation of public resources. 
From a public health perspective, insights from potential landscapes can help policymakers develop targeted interventions during pandemics and other crises.
Additionally, the potential landscape approach offers decision-makers a dynamic tool for tracking mobility trends, enabling them to adjust infrastructure and regulations in response to evolving urban needs.
Future studies with higher temporal resolution and real-time data will be crucial for expanding the scope of these applications and enhancing the precision of mobility analyses.


\section*{Statements and Declarations}
\subsection*{Competing Interests} 
The authors report no conflicts of interest.

\section*{Data Availability Statement}
The mobility flow data for Japan utilized in this study are available for purchase from LocationMind Inc., a Japanese company \cite{LocationMind} (for inquiries, use their contact form \cite{LocationMind_contact}). The data product is named ``LocationMind xPop.''

\section*{Acknowledgements}
Y.D. is supported by JST SPRING, Grant Number JPMJSP2114. 
T.A. is supported by JSPS KAKENHI Grant Number JP24K07699, and JSPS KAKENHI Grant Number JP24H01473.
N.F. is supported by JSPS KAKENHI Grant Number JP24K03007, and
JST PRESTO Grant Number JPMJPR21RA, Japan.
This work also receives support from the Research Institute for Mathematical Sciences, an International Joint Usage/Research Center located in Kyoto University.


\section*{Supplementary Notes}

\subsection*{Supplementary Note 1: Non-zero mobility pairs}
\begin{figure}[h!] 
  \centering
  \includegraphics[width=\textwidth]{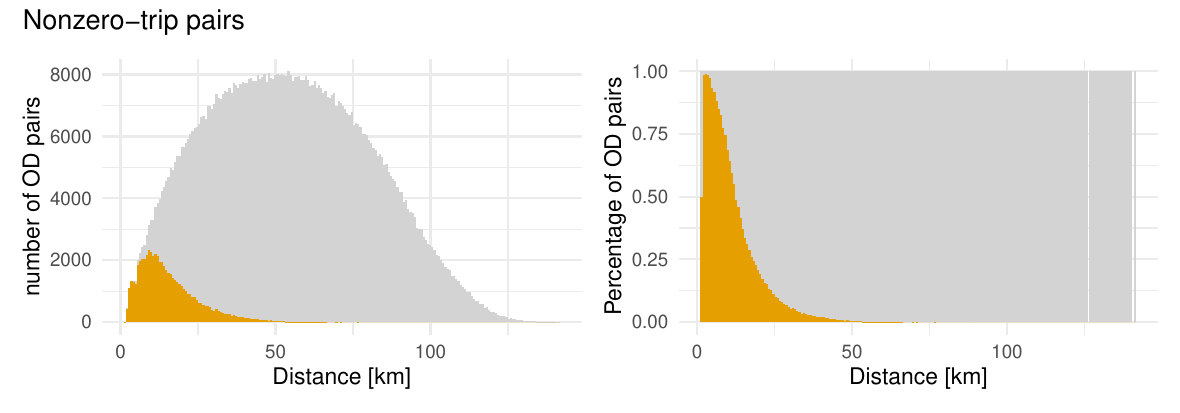}
  \caption{Percentage of OD pairs $(i,j)$ with non-zero mobility as a function of spatial distance $d_{ij}$. In the range of distance, number (left) and percentage (right) of non-zero mobility pairs are plotted.}
  \label{fig_s1}
\end{figure}

To investigate the dependence of mobility flow on distance, Figure~\ref{fig_s1} shows the percentage of non-zero mobility pairs ($M_{ij} > 0$ or $M_{ji} > 0$) as a function of road distance in the Japanese mobility dataset \cite{LocationMind}. A rapid decline is observed as the distance increases. Notably, even at a relatively short distance of approximately $d_{ij} \sim 40$ km, around half of the mobility pairs exhibit zero flow volume. 

\subsection*{Supplementary Note 2: Sensitivity analysis by testing different values of $\theta$} 
In the first reduction step, we applied a weighting method with a threshold distance $\theta$ \cite{AFF2023}:
\begin{align*}
  W_{ij} = \begin{cases}
    1 \quad  &\text{mobility pair $\{ (i,j) \mid d_{ij} \le \theta\}$} \\
    0 \quad & \text{mobility pair $\{ (i,j) \mid d_{ij} > \theta\}$}
\end{cases}.
\end{align*}
The threshold $\theta$ is the distance below which the majority of trips occur, based on the probabilistic criterion $p(X < \theta) = 0.99$.
This criterion was chosen to include the majority of trips, while excluding possible outliers.

Table \ref{tbl:sensitivity} shows a sensitivity analysis by testing different criteria ($p$ = 0.95 and 0.9), which led to different thresholds $\theta$.
These results demonstrate that the resulting potential landscapes with modified parameters were very similar, maintaining high correlations even when the threshold $\theta$ was changed significantly.

\begin{table}[!h]
\centering
\caption{Correlation between the generated potential landscapes with modified $p$.}
\centering
\begin{tabular}[t]{rlllrrlr}
\toprule
Year & Condition & $p$ & $p$ (modified) & $\theta$ & $\theta$ (modified) & $\theta$ (changed) & Correlaton\\
\midrule
2019 & Weekday & 0.99 & 0.95 & 32.54 & 19.79 & -39\% & 0.92\\
2019 & Holiday & 0.99 & 0.95 & 22.58 & 12.62 & -44\% & 0.83\\
2021 & Weekday & 0.99 & 0.95 & 30.50 & 18.05 & -41\% & 0.92\\
2021 & Holiday & 0.99 & 0.95 & 19.36 & 11.53 & -40\% & 0.90\\
\addlinespace
2019 & Weekday & 0.99 & 0.9 & 32.54 & 14.31 & -56\% & 0.87\\
2019 & Holiday & 0.99 & 0.9 & 22.58 & 9.09 & -60\% & 0.89\\
2021 & Weekday & 0.99 & 0.9 & 30.50 & 13.08 & -57\% & 0.87\\
2021 & Holiday & 0.99 & 0.9 & 19.36 & 8.68 & -55\% & 0.76\\
\bottomrule
\end{tabular}
  \label{tbl:sensitivity}
\end{table}

\subsection*{Supplementary Note 3: Variances explained by principal components}
\begin{figure}[h!] 
  \centering
  \includegraphics[width=0.5\textwidth]{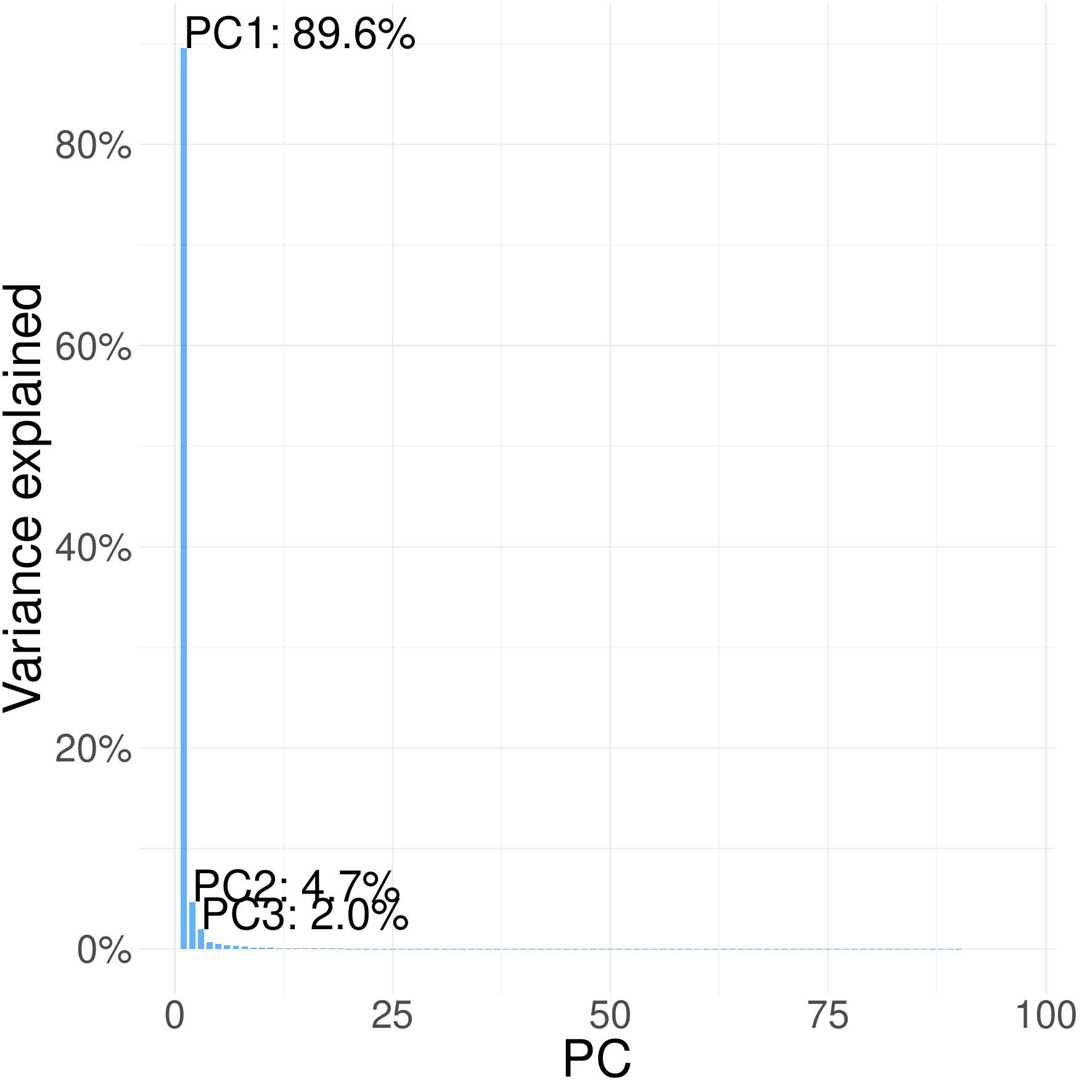}
  \caption{Screeplot of the PCA results for the combined data from weekdays and holidays in 2019 and 2021. The first three principal components capture the majority of the variance.}
  \label{fig_s3}
\end{figure}

We conducted a PCA on the combined data from weekdays and holidays in both 2019 and 2021. Notably, the first three principal components capture the majority of the variance, as shown in Fig.~\ref{fig_s3}. This indicates that they play a crucial role in explaining the underlying mobility patterns.
\clearpage

\subsection*{Supplementary Note 4: Time evolution of potential landscape}
\begin{figure}[h!] 
  \centering
\subcaptionbox{00:00 - 01:00}[0.24\linewidth]{ \includegraphics[width=\linewidth]{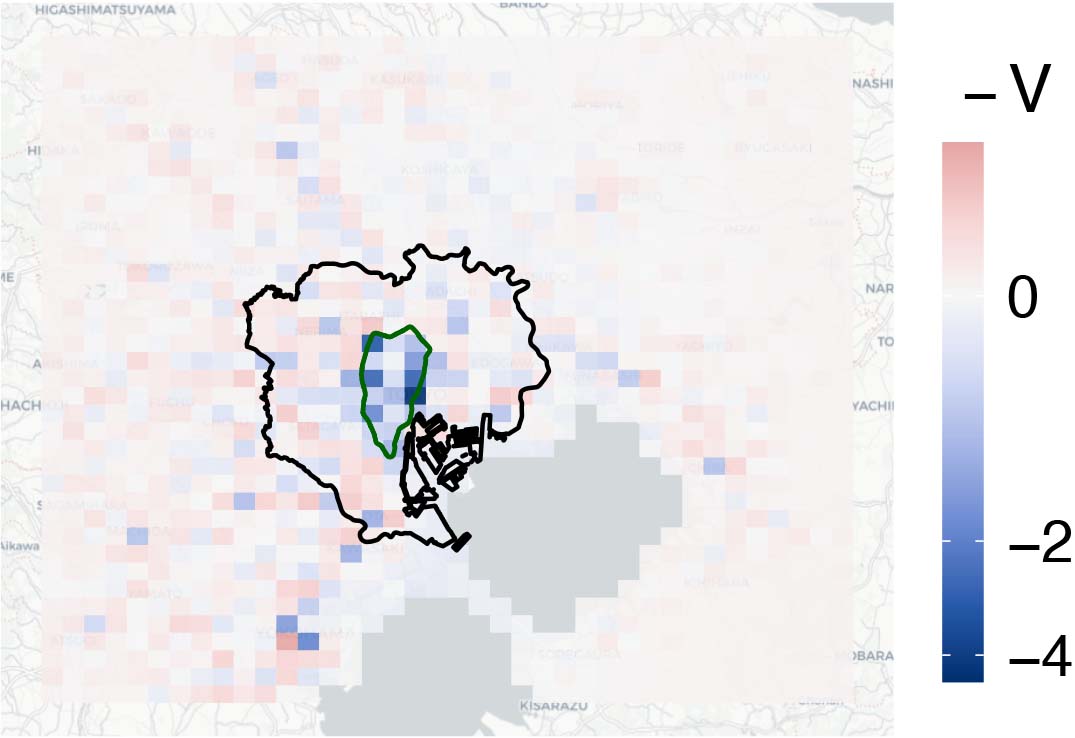} }
\subcaptionbox{00:00 - 01:00}[0.24\linewidth]{ \includegraphics[width=\linewidth]{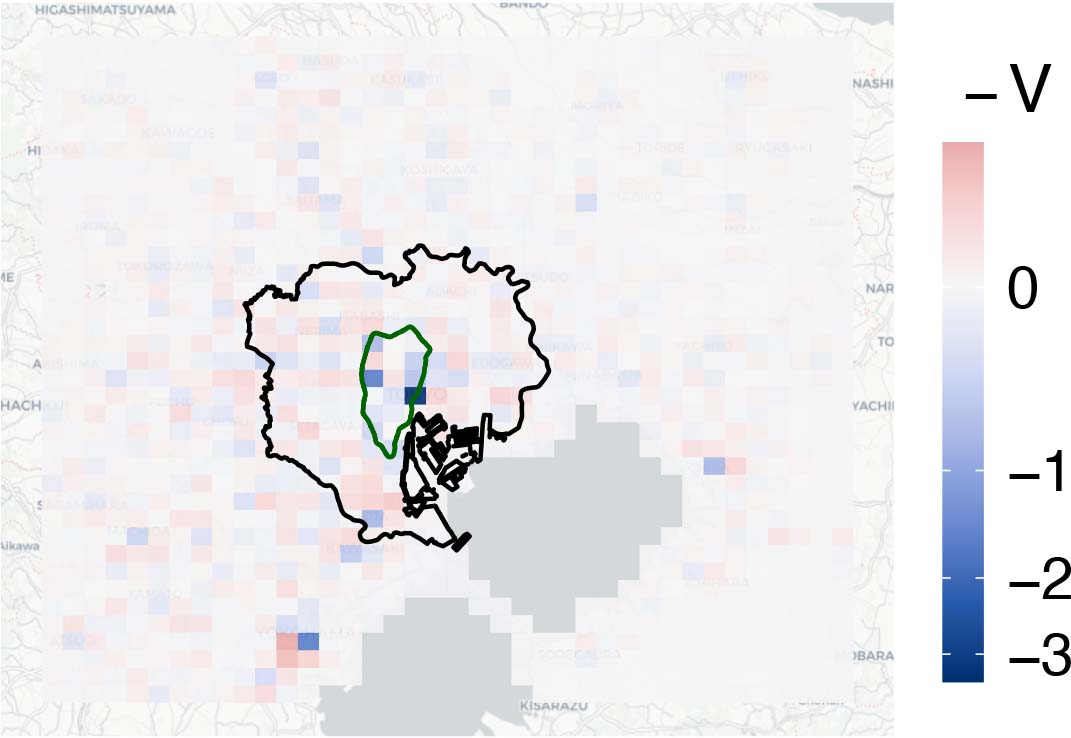} }
\subcaptionbox{01:00 - 02:00}[0.24\linewidth]{ \includegraphics[width=\linewidth]{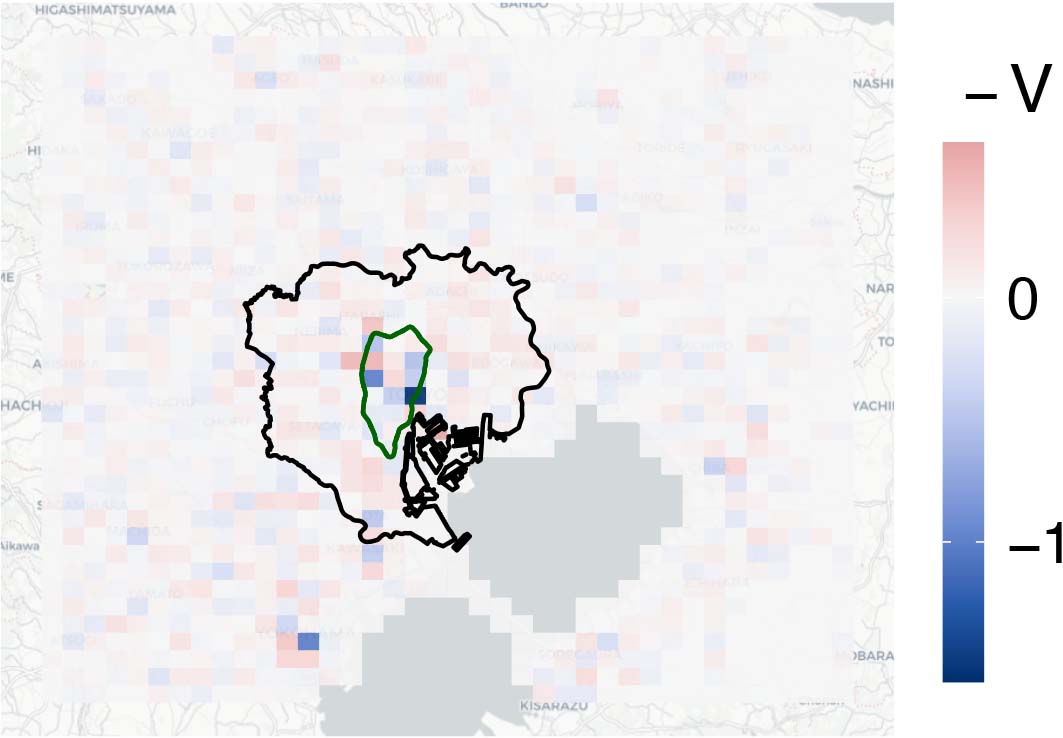} }
\subcaptionbox{02:00 - 03:00}[0.24\linewidth]{ \includegraphics[width=\linewidth]{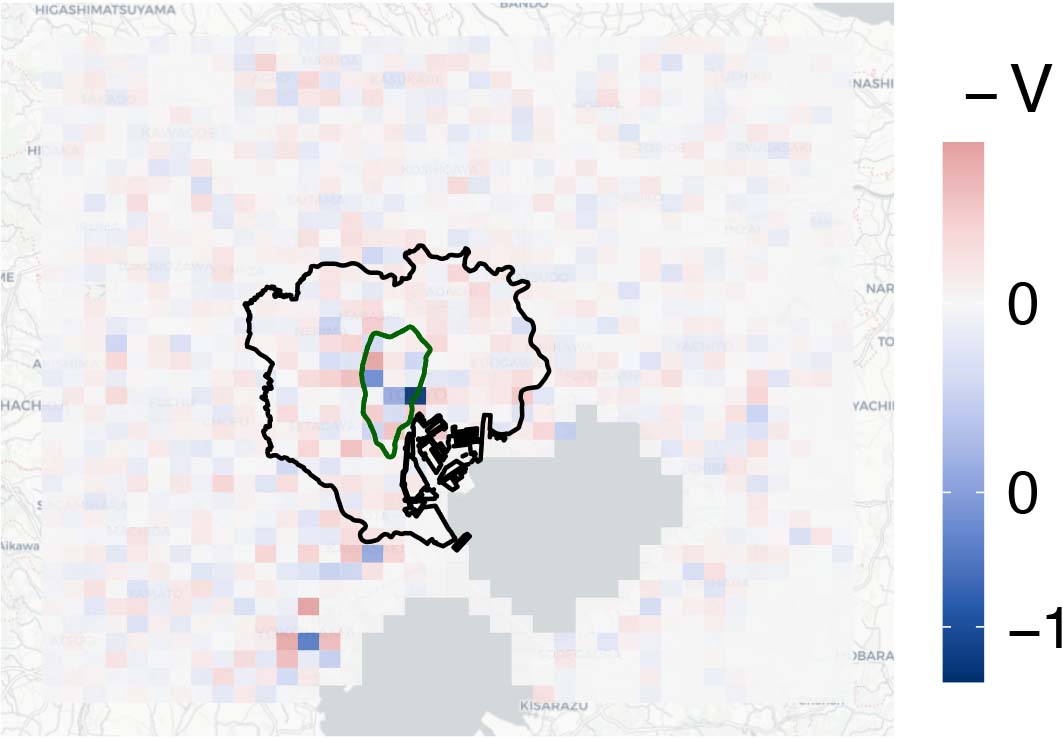} }
\subcaptionbox{03:00 - 04:00}[0.24\linewidth]{ \includegraphics[width=\linewidth]{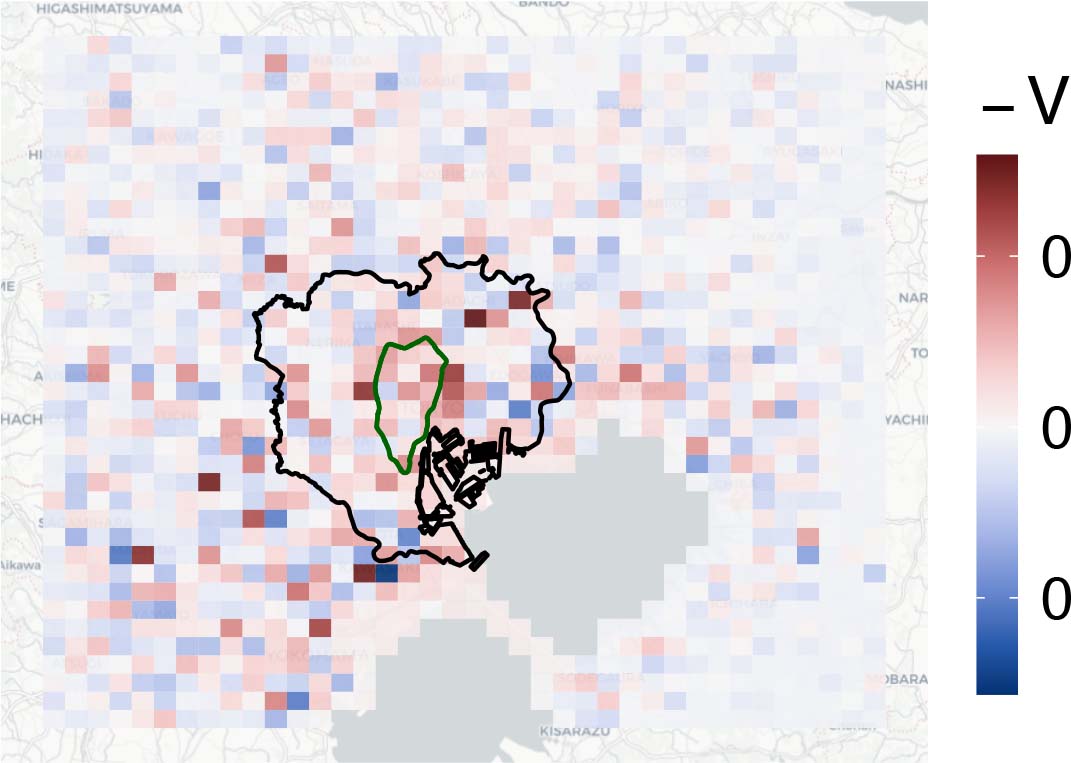} }
\subcaptionbox{04:00 - 05:00}[0.24\linewidth]{ \includegraphics[width=\linewidth]{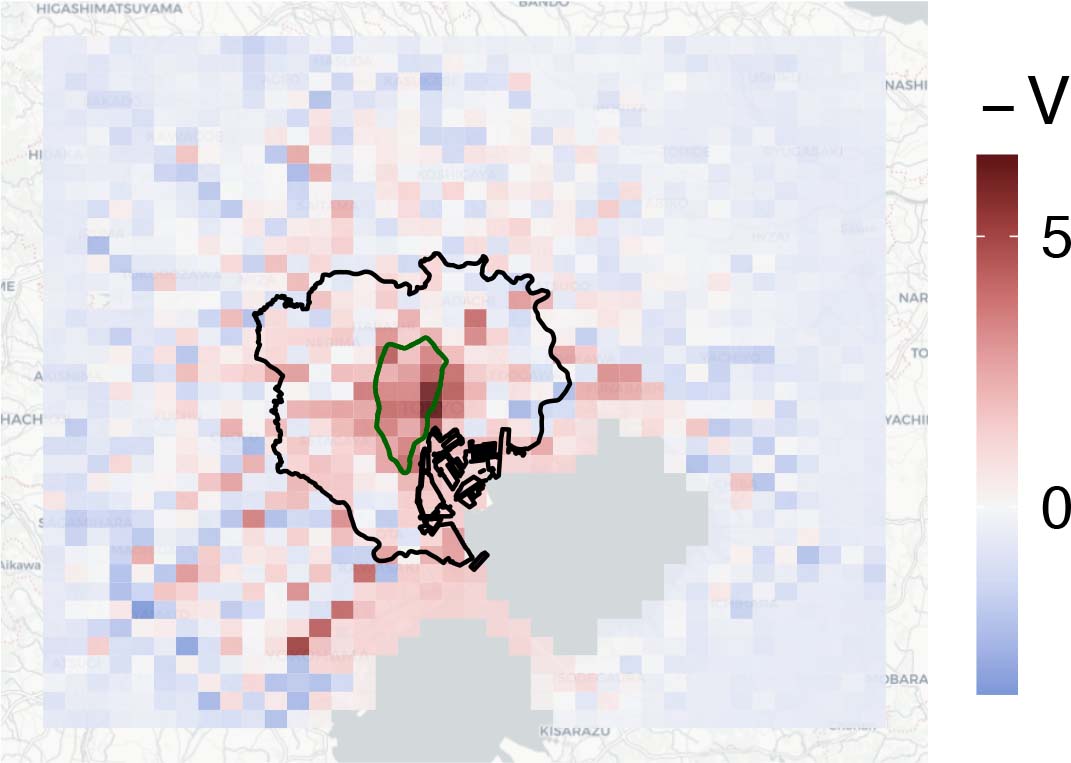} }
\subcaptionbox{05:00 - 06:00}[0.24\linewidth]{ \includegraphics[width=\linewidth]{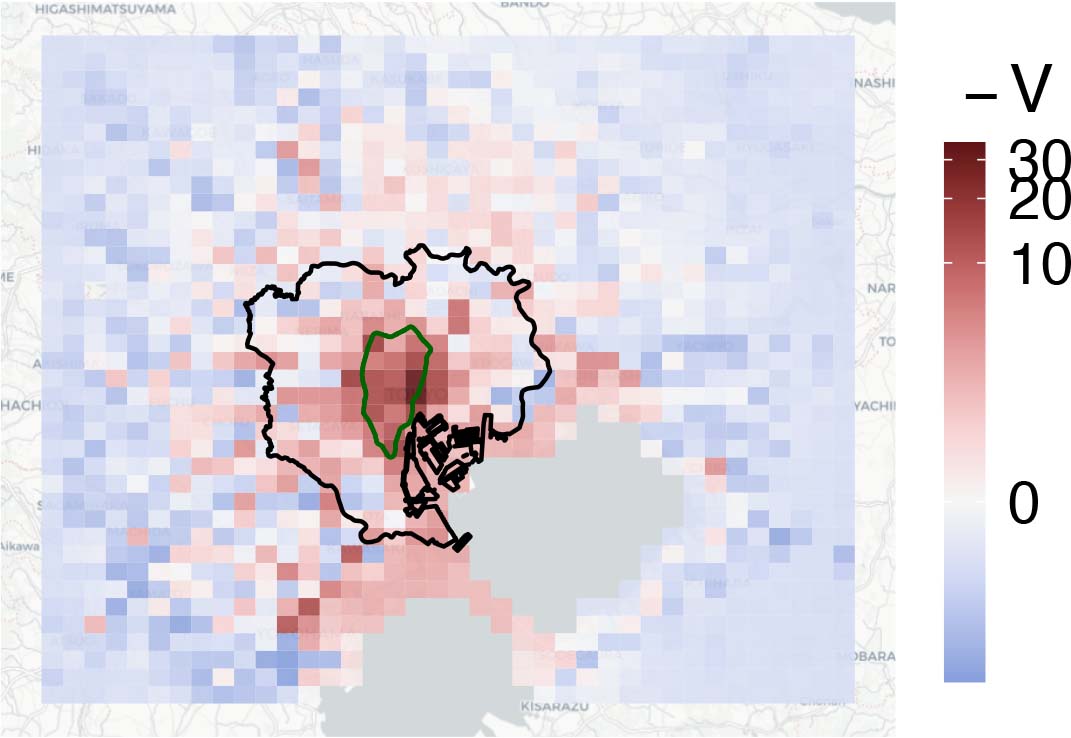} }
\subcaptionbox{06:00 - 07:00}[0.24\linewidth]{ \includegraphics[width=\linewidth]{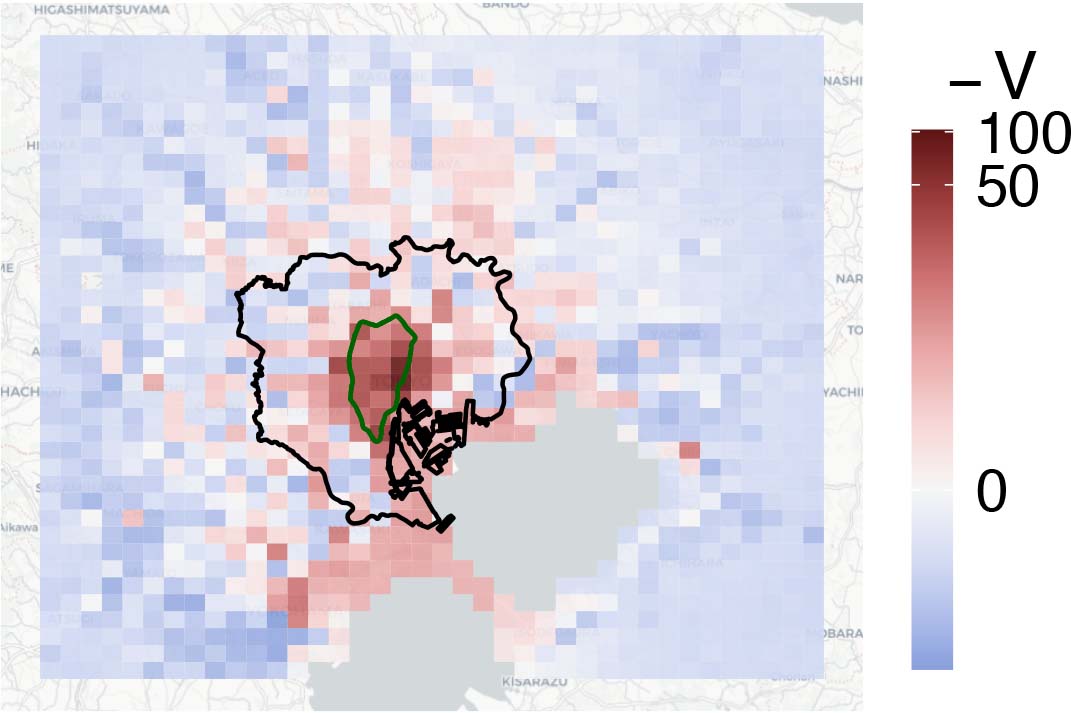} }
\subcaptionbox{07:00 - 08:00}[0.24\linewidth]{ \includegraphics[width=\linewidth]{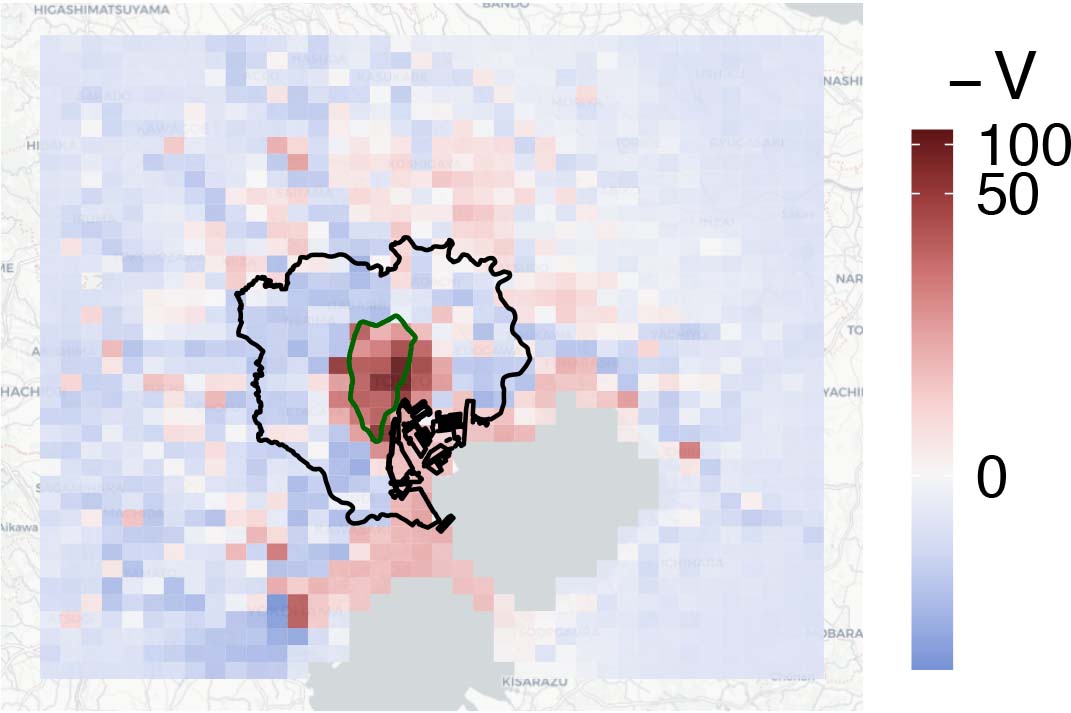} }
\subcaptionbox{08:00 - 09:00}[0.24\linewidth]{ \includegraphics[width=\linewidth]{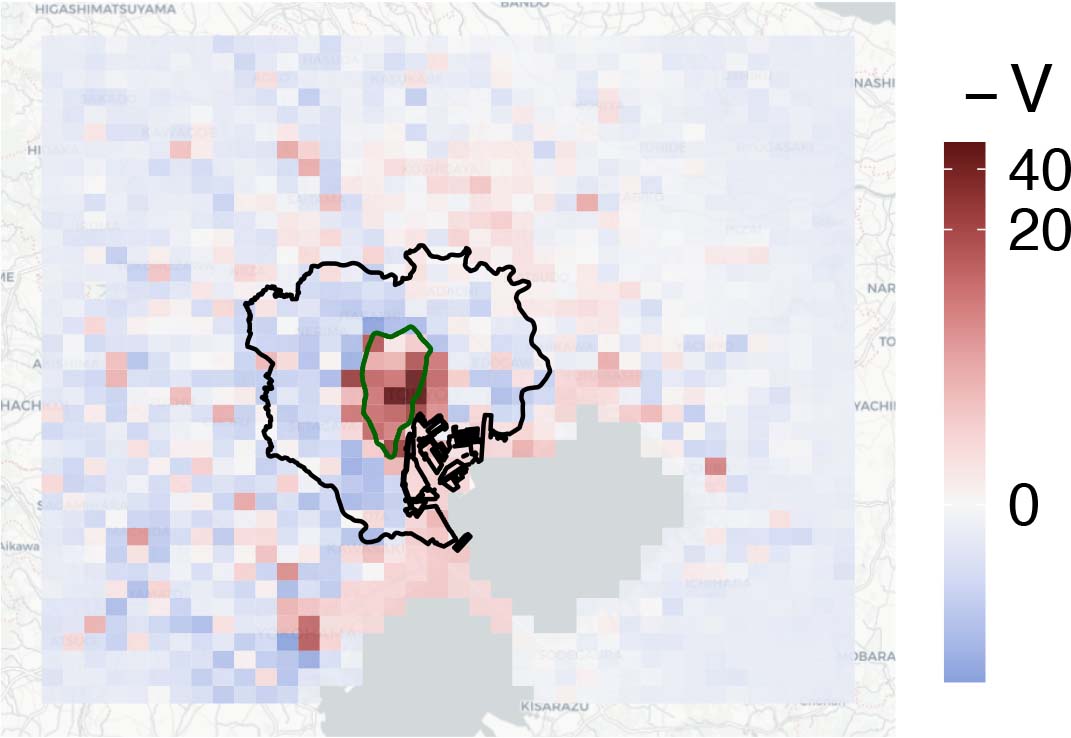} }
\subcaptionbox{09:00 - 10:00}[0.24\linewidth]{ \includegraphics[width=\linewidth]{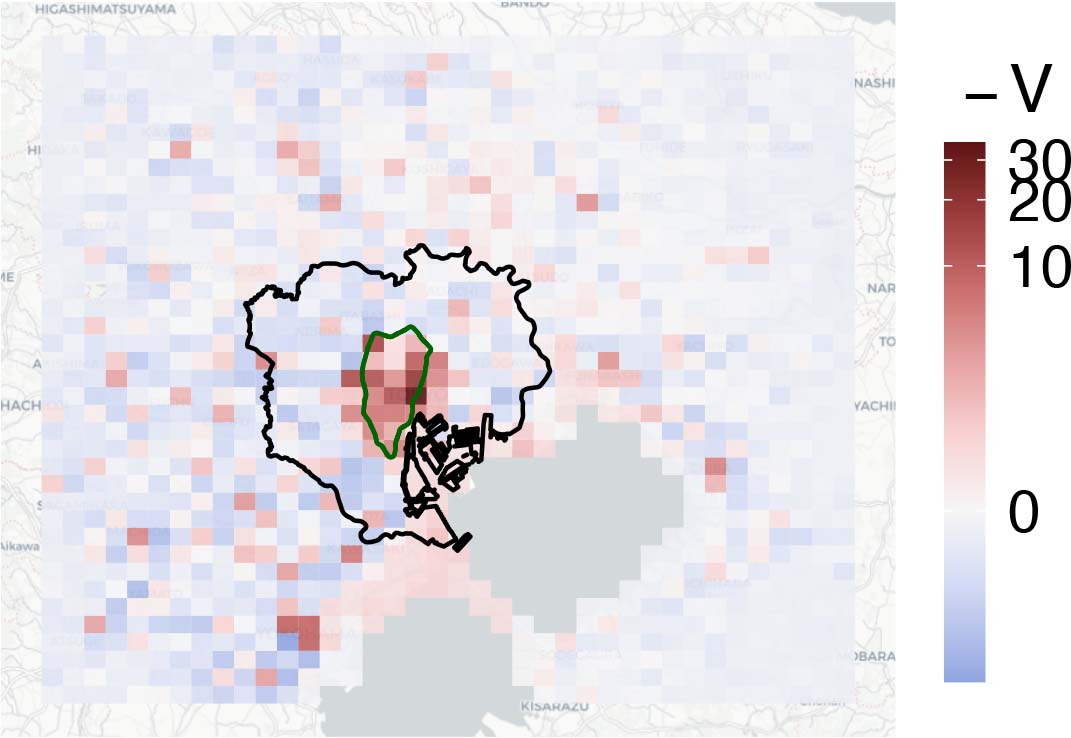} }
\subcaptionbox{10:00 - 11:00}[0.24\linewidth]{ \includegraphics[width=\linewidth]{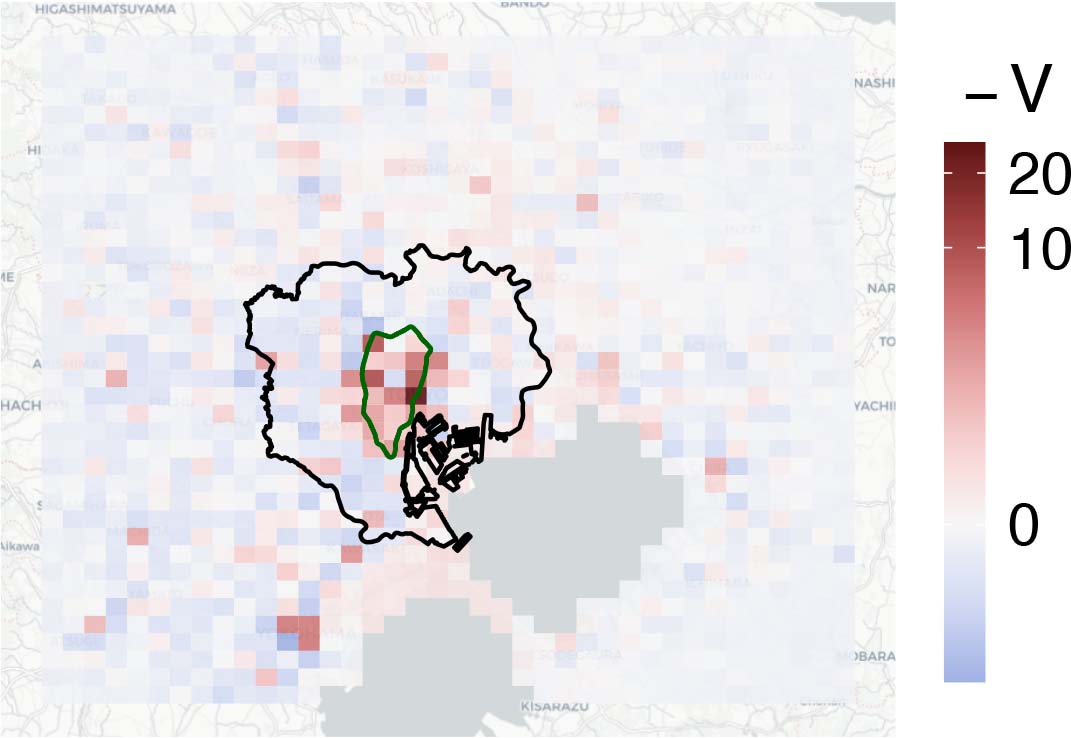} }
  \caption{Time evolution of potential landscape at different hours of the day in 2019 weekday.}
  \label{fig_potential_1}
\end{figure}

\begin{figure}[h!] 
  \centering
\subcaptionbox{11:00 - 12:00}[0.24\linewidth]{ \includegraphics[width=\linewidth]{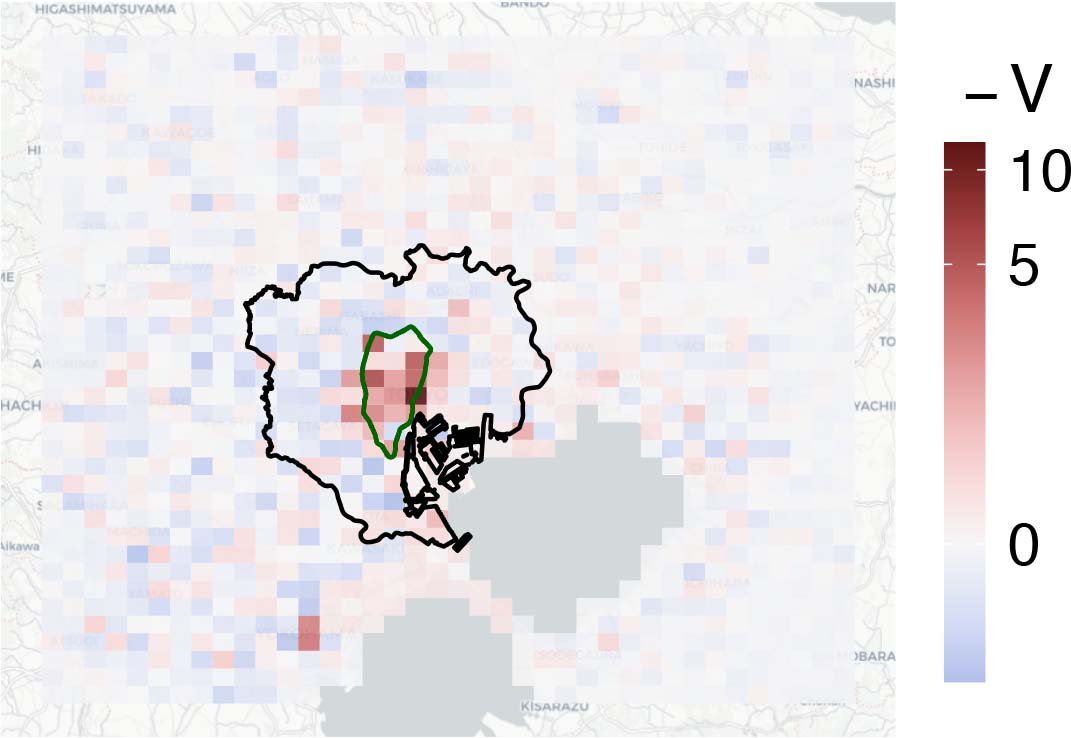} }
\subcaptionbox{12:00 - 13:00}[0.24\linewidth]{ \includegraphics[width=\linewidth]{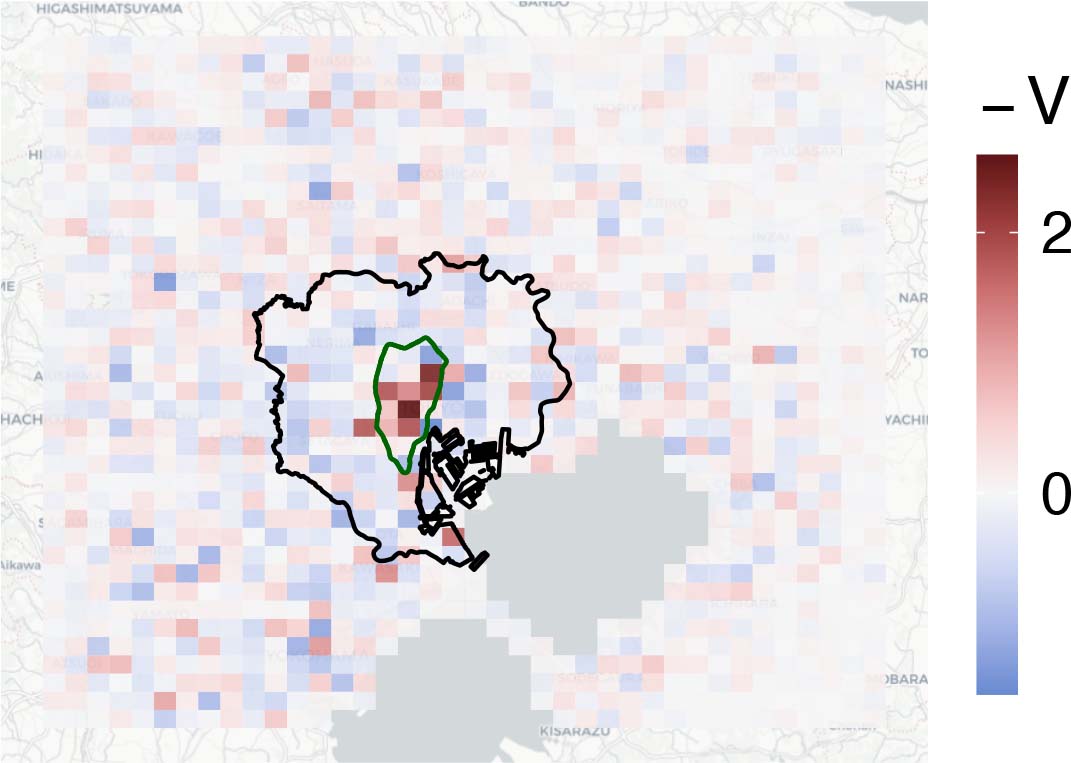} }
\subcaptionbox{13:00 - 14:00}[0.24\linewidth]{ \includegraphics[width=\linewidth]{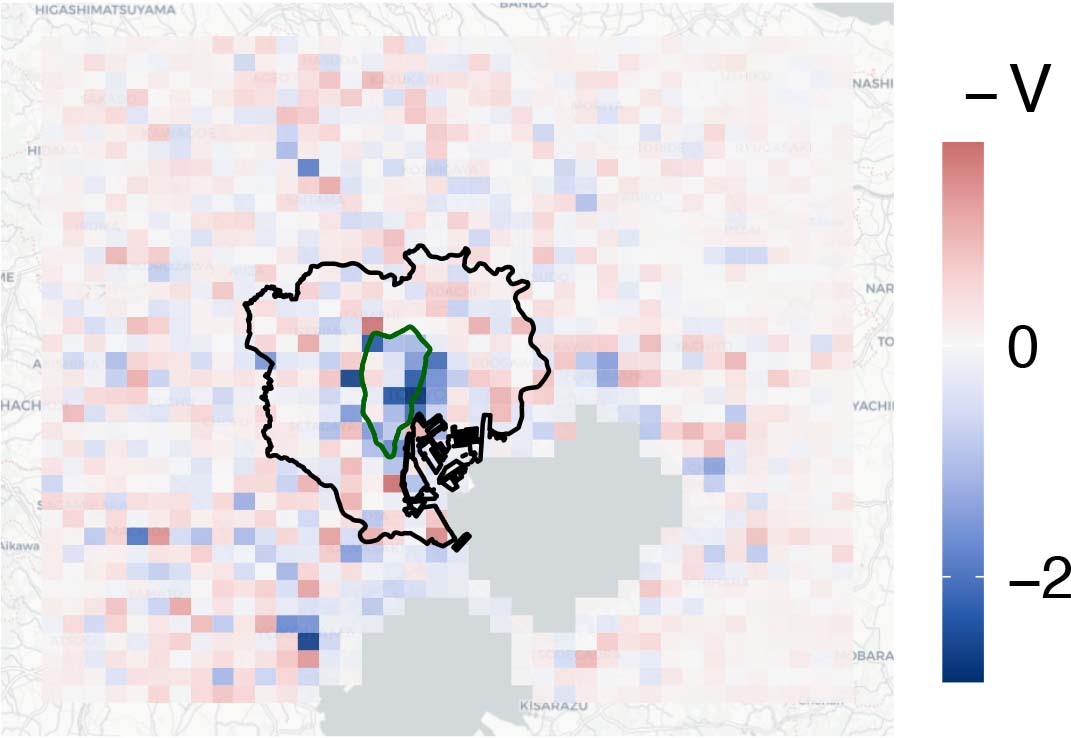} }
\subcaptionbox{14:00 - 15:00}[0.24\linewidth]{ \includegraphics[width=\linewidth]{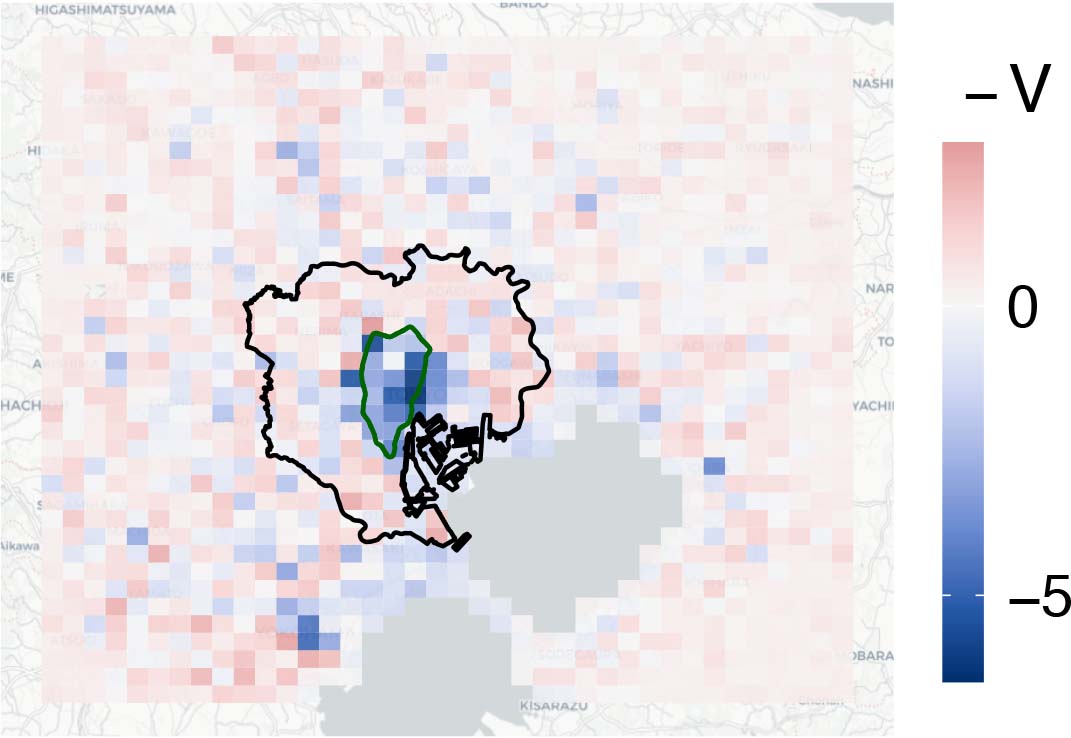} }
\subcaptionbox{15:00 - 16:00}[0.24\linewidth]{ \includegraphics[width=\linewidth]{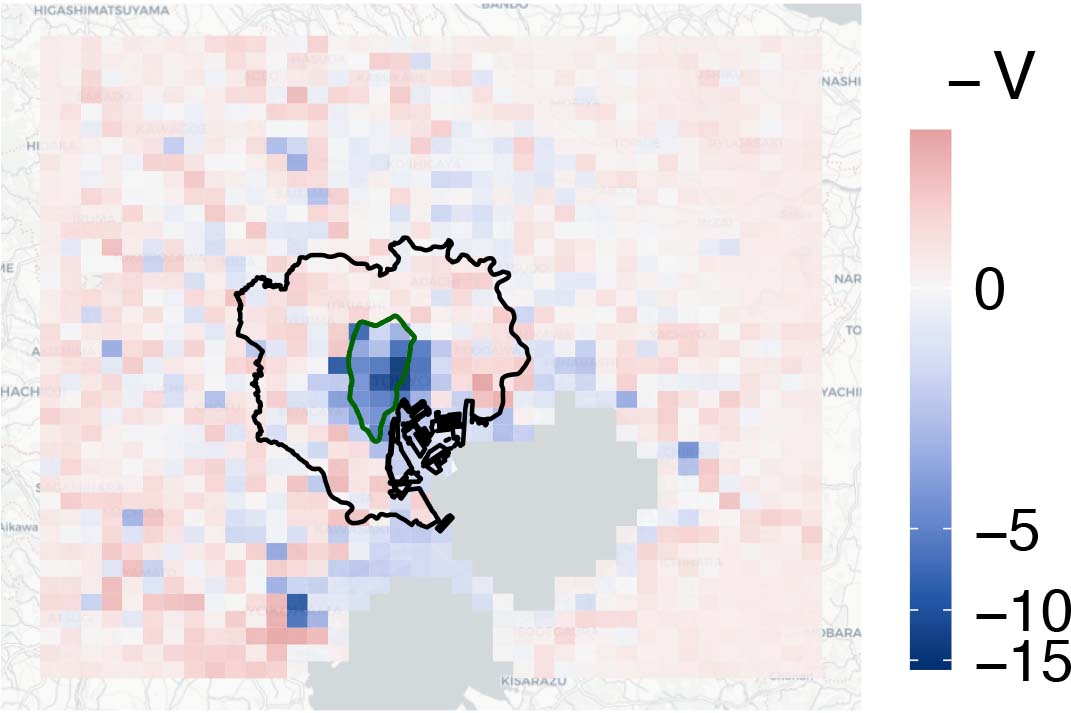} }
\subcaptionbox{16:00 - 17:00}[0.24\linewidth]{ \includegraphics[width=\linewidth]{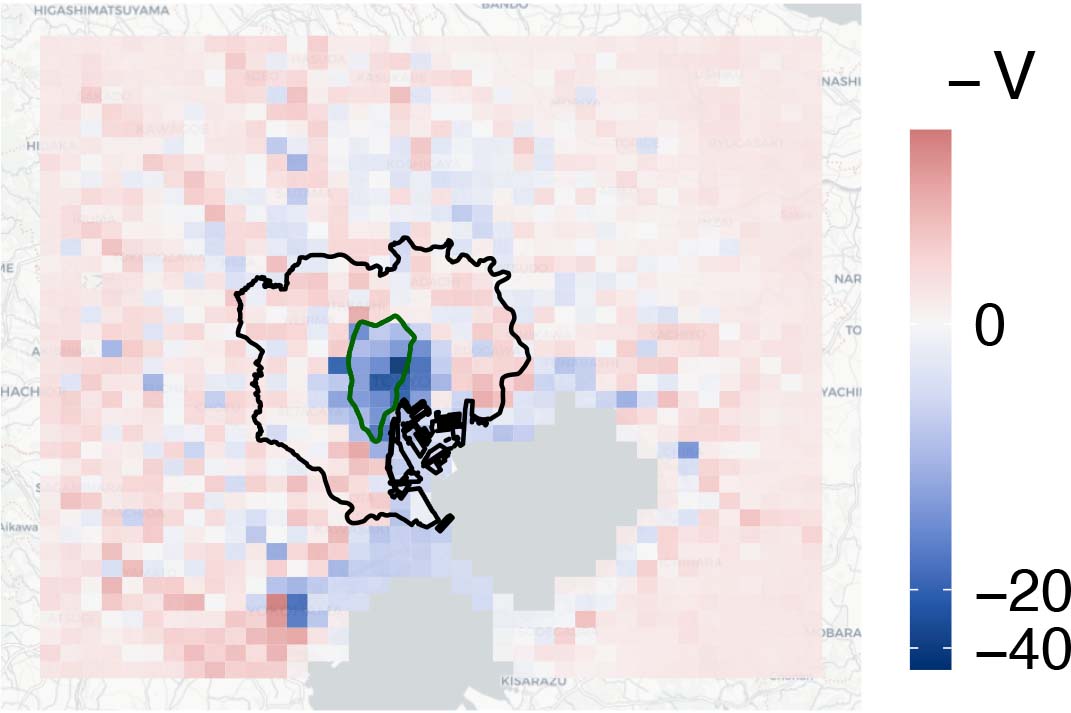} }
\subcaptionbox{17:00 - 18:00}[0.24\linewidth]{ \includegraphics[width=\linewidth]{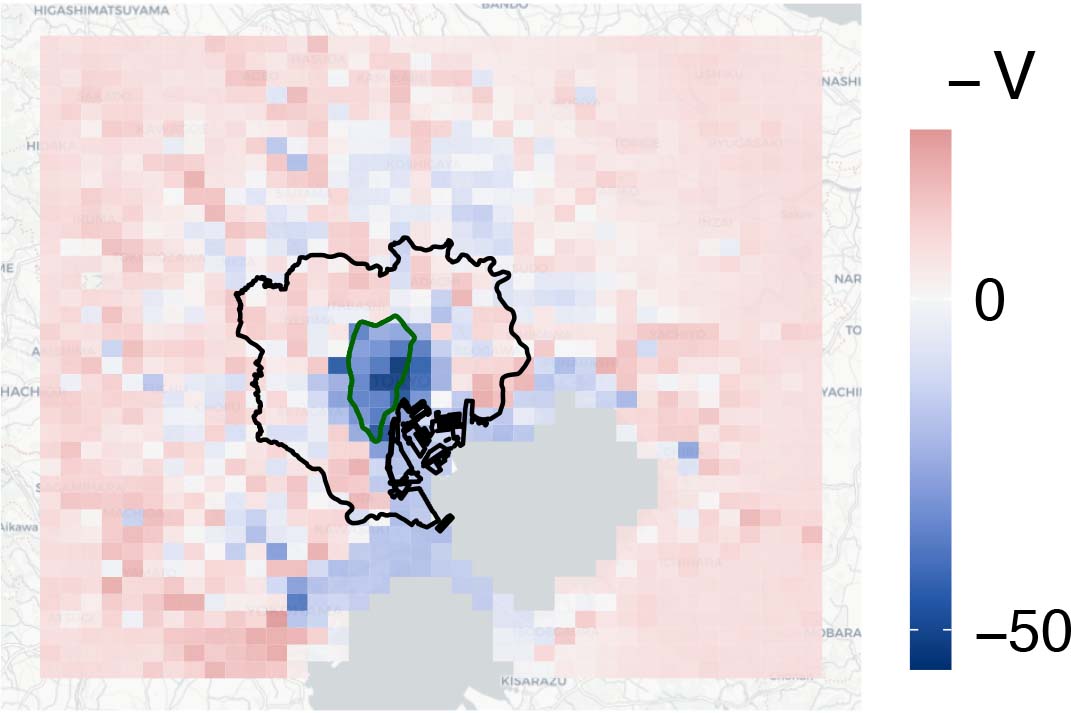} }
\subcaptionbox{18:00 - 19:00}[0.24\linewidth]{ \includegraphics[width=\linewidth]{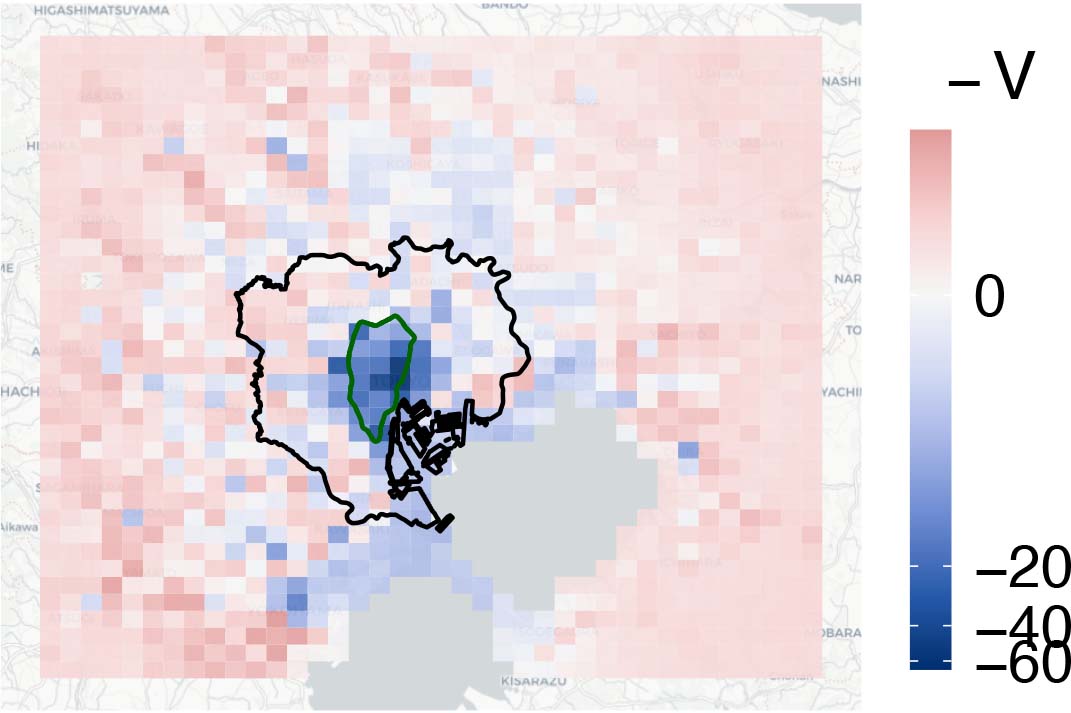} }
\subcaptionbox{19:00 - 20:00}[0.24\linewidth]{ \includegraphics[width=\linewidth]{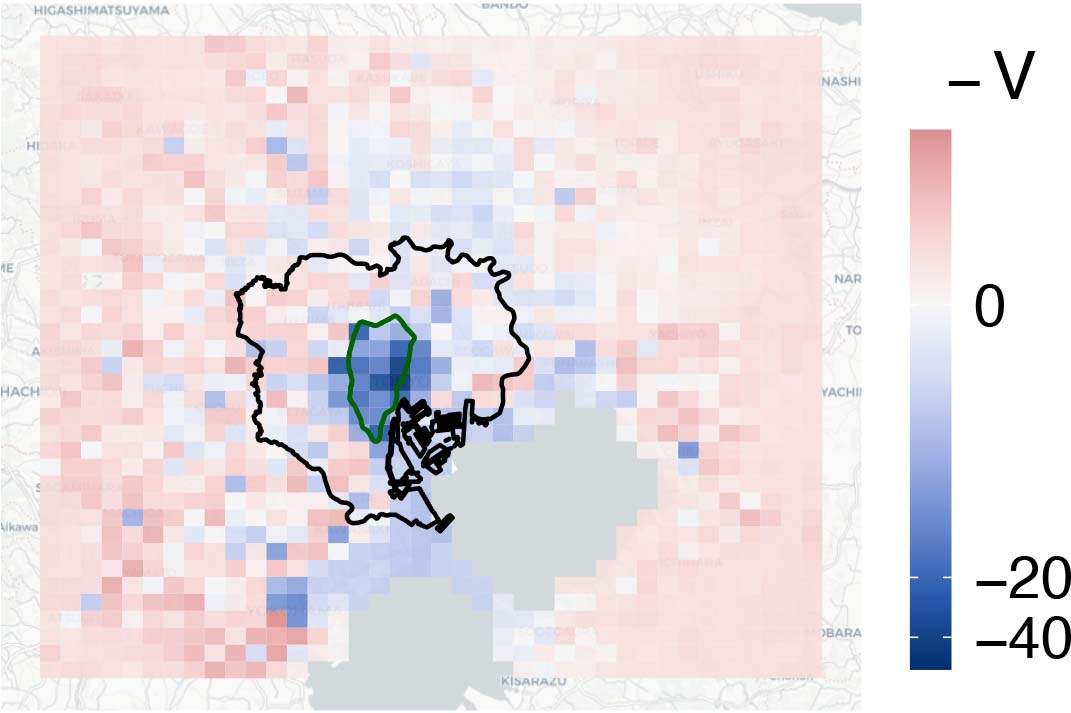} }
\subcaptionbox{20:00 - 21:00}[0.24\linewidth]{ \includegraphics[width=\linewidth]{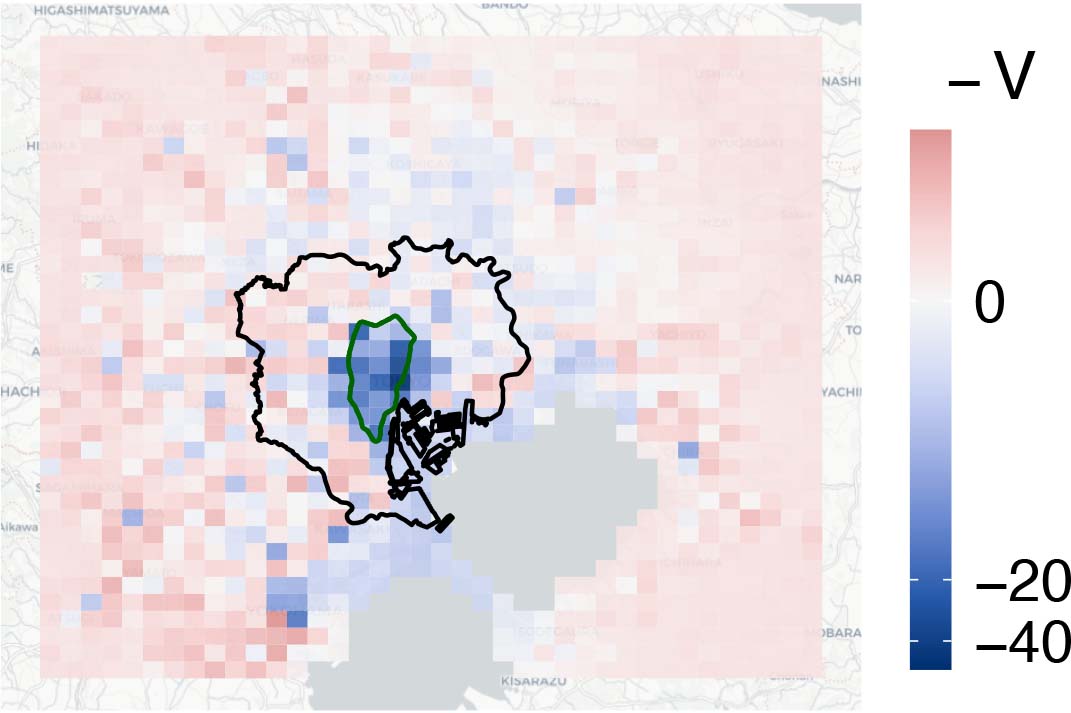} }
\subcaptionbox{21:00 - 22:00}[0.24\linewidth]{ \includegraphics[width=\linewidth]{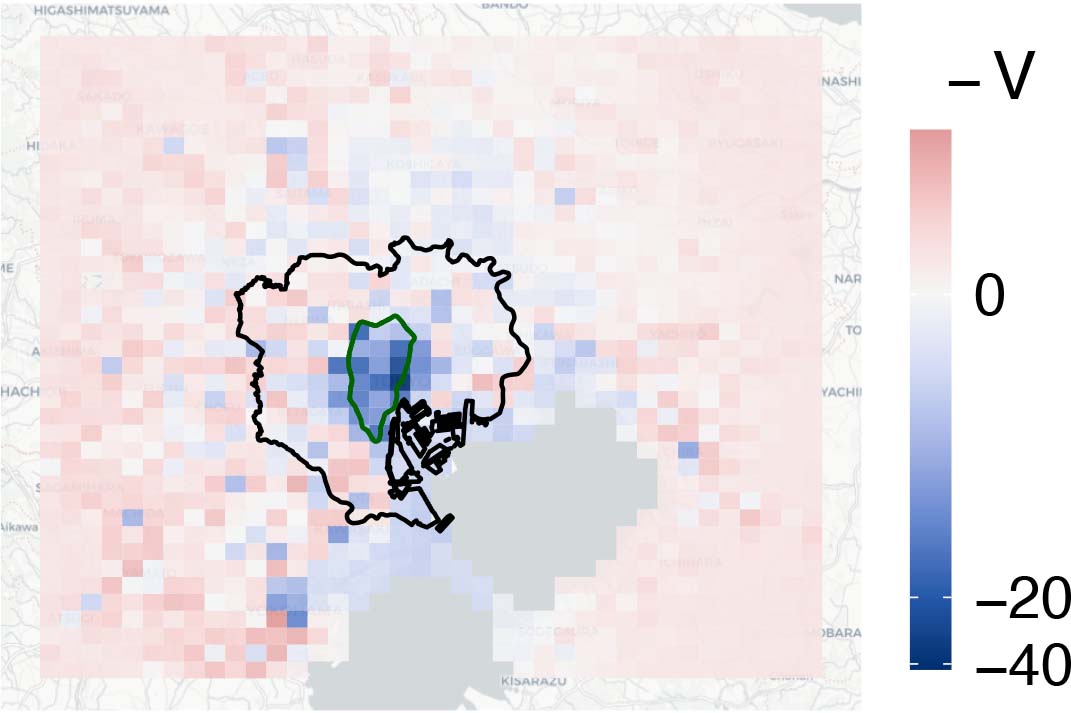} }
\subcaptionbox{22:00 - 23:00}[0.24\linewidth]{ \includegraphics[width=\linewidth]{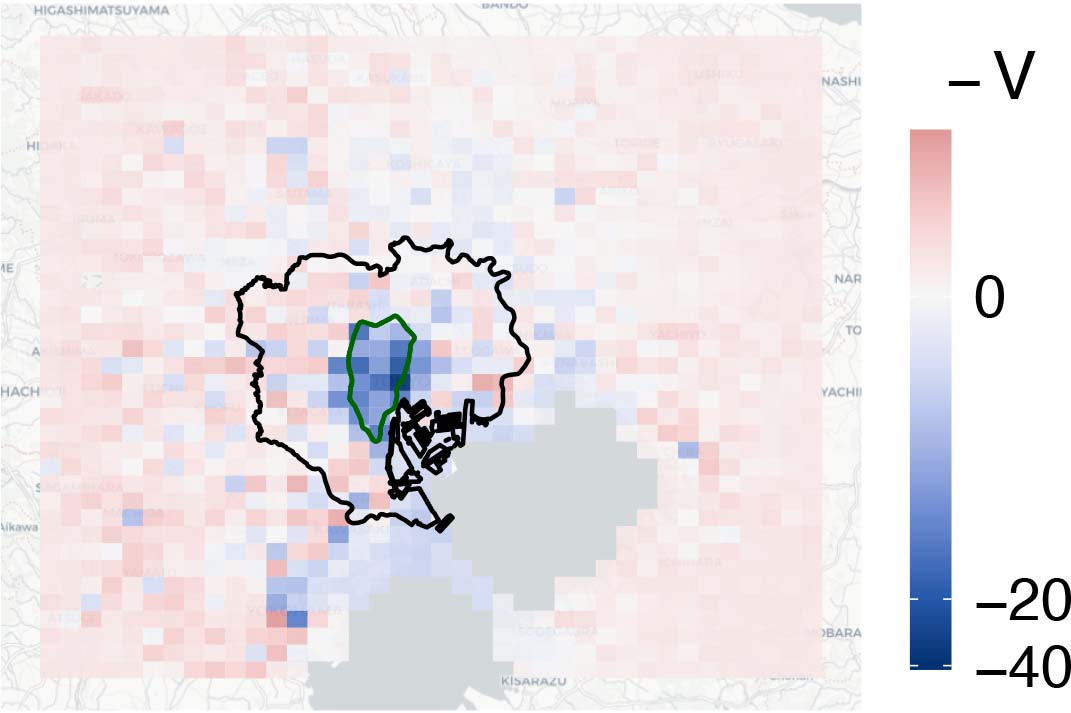} }
  \caption{Time evolution of potential landscape at different hours of the day in 2019 weekday.}
  \label{fig_potential_2}
\end{figure}

The hourly detailed potential landscapes for the 2019 weekday scenario are shown in Figs.~\ref{fig_potential_1} and \ref{fig_potential_2}. The potential landscapes exhibit a clear pattern as discussed in the main text.

\clearpage

\subsection*{Appendix: Data and map sources for Supplementary Information figures}
The background map layouts in Figs.~\ref{fig_potential_1} and \ref{fig_potential_2} are based on data by \textcopyright{} OpenStreetMap contributors, available under the Open Database License (ODbL). Map tiles by \textcopyright{} CARTO. The results in Figs.~\ref{fig_s1}, \ref{fig_s3}, \ref{fig_potential_1}, and \ref{fig_potential_2} were obtained by using data from ``LocationMind xPop \textcopyright{} LocationMind Inc'' \cite{LocationMind}.

\clearpage

\bibliography{ref}

\end{document}